\documentclass[12pt]{article}

\usepackage{amsfonts}
\usepackage{amssymb}
\usepackage{graphicx}
\usepackage{amsmath}
\usepackage{makeidx}
\usepackage[T1]{fontenc}
\usepackage{color}
\usepackage{graphicx}
\usepackage{makeidx}
\usepackage{subcaption}
\usepackage{float}
\usepackage[section]{placeins}

\newtheorem{theorem}{Theorem}

\newtheorem{adefinition}[theorem]{Definition}

\newtheorem{aexample}[theorem]{Example}

\newtheorem{aproblem}[theorem]{Problem}

\newtheorem{acomment}[theorem]{Comment}

\newtheorem{aremark}[theorem]{Remark}

\numberwithin{equation}{section} \numberwithin{theorem}{section}

\makeatother
\makeindex

\begin{document}

\title{The Dynamics of Quantum Correlations \\ of Two Qubits
in a Common
Environment \\ Modeled by Large Random Matrices}
\author{ E.Bratus and L.Pastur \\
\small B. Verkin Institute for Low Temperatures Physics \\
\small and Engineering, Kharkiv, Ukraine }

\maketitle

\begin{abstract}
This paper is a continuation of our previous paper
\cite{Br-Pa:18}, in which we have studied the dynamics of quantum
correlations of two qubits embedded each into its own disordered
multiconnected environment. We modeled the environment by random
matrices of large size allowing for a possibility to describe
meso- and even nanoenvironments. In this paper we also study the
dynamics of quantum correlations of two qubits but embedded into a
common environment which we also model by random matrices of large
size. We obtain the large size limit of the reduced density matrix
of two qubits. We then use an analog of the Bogolyubov-van Hove (also known as the Born-Markov)
approximation of the theory of open systems and statistical
mechanics. The approximation does not imply in general the
Markovian evolution in our model but allows for  sufficiently
detailed analysis both analytical and numerical of the evolution
of several widely used quantifiers of quantum correlation, mainly
entanglement. We find a number of new patterns of qubits dynamics
comparing with the case of independent environments studied in
\cite{Br-Pa:18} and displaying the role of dynamical (indirect,
via the environment) correlations in the enhancing and
diversification of qubit evolution. Our results,
(announced in \cite{Br-Pa:20}), can be viewed as a
manifestation of the universality
of certain properties of the decoherent qubit evolution which have
been found previously in various exact and approximate versions of
two-qubit models with macroscopic bosonic environment.
\end{abstract}


\section{Introduction}

Entanglement is a counterintuitive and an intrinsically quantum form of
correlations between the parts of quantum systems, whose state cannot be
written as the product of states of the parts. It is a basic ingredient of
the quantum theory, having a great potential for applications in
quantum technology \cite{Ho-Co:09,Ni-Ch:00,Oh-Vo:11}. Inevitable
interactions of quantum systems with an environment degrade in general
quantum correlations, entanglement in particular. This is why the studies of
dynamical aspects of entanglement, including entanglement behavior under
interaction with the environment, are of great interest and importance for
quantum information theory. They also make the link of the field with the
fundamental problems of quantum dynamics, in particular, those of the theory
of open systems and statistical mechanics \cite{Br-Pe:07,Da:76,de-Al:17,Di-Co:98,We:08}.
In view of this the models of dynamics of qubits, the basic entities of quantum
information science, embedded in an environment comprise an active branch of
quantum information theory and adjacent fields, see \cite{Ao-Co:15,Br-Co:15,Lo-Co:13,Ri-Co:14,Yu-Eb:09} for
reviews. In particular, there exists a certain amount of models, where the
qubits-environment Hamiltonians include random matrices of large size,
see our paper
\cite{Br-Pa:18} for the comparative analysis of these models. It is
worth mentioning that random matrices have been widely using to describe
complex quantum systems of large but not necessary macroscopic size, see
e.g., \cite{Ak-Co:11,Fo:10,Pa-Sh:11} for results and references. In particular, in our recent
work \cite{Br-Pa:18} we analyzed the evolution of two qubits
interacting with

(i) either a two-component environment with dynamically independent
components each interacting with its "own" qubit,

(ii) or a one-component environment interacting with one of two qubits while
the second qubit is free (so called ancilla).

In both cases the dynamics of the whole system is the tensor product of
dynamics of its two parties (one qubit plus its environment if any). This
allowed us to use the results on a random matrix
model of the one qubit dynamics given in \cite{Le-Pa:03} and to study
a number of properties of the evolution of quantum correlations, entanglement in particular,
including the properties
found earlier for other models of
environment, mostly for the free boson environment and its various
approximate versions \cite{Ad-Co:13,Da-Co:12,Lo-Co:13,Yu-Eb:04,Yu-Eb:09,Zy-Co:01}.

In this paper we will present our
results on a physically different and, we believe, quite interesting model
of two qubits interacting with a common environment also modeled by random
matrices. In this case we have to work out the corresponding dynamics anew
by using an extension of random matrix techniques of our earlier works \cite%
{Le-Pa:03,Br-Pa:18,Br-Pa:20,Pa-Sh:11}. As a result, we are able to study
a variety of interesting time evolutions both new and found
in other models of environment for the widely used quantifiers of
quantum correlations (the concurrence, the negativity, the quantum
discord and the von Neumann entropy).

The paper is organized as follows. In Section 2 we describe our
model and the characteristics (quantifiers) of quantum
correlations to be studied. In Section 3 we present our both
analytical and numerical results obtained in the framework of the
model. Section 4 contains the proof of the basic formulas for
the large size limit of the reduced density matrix which
have been announced in \cite{ Br-Pa:20}.
To make our presentation sufficiently selfconsistent and not
too long, we use certain results and outline certain reasonings
of our earlier works \cite{Le-Pa:03,Br-Pa:18,Br-Pa:20}. Thus,
the paper is partly a self review.

\section{Model}

\subsection{Generalities}

We will use the general setting that has been worked out in a number of
works on the dynamics of qubits embedded in a sufficiently "large"
environment, see e.g. \cite{Ao-Co:15,Da-Co:12,Ho-Co:09,Lo-Co:13,Yu-Eb:09}
and earlier in the theory of open systems \cite{Br-Pe:07,Da:76,de-Al:17,Di-Co:98,We:08}.

The basic quantity to be studied here is the reduced density matrix of
qubits defined as follows. Let
\begin{equation}
\rho _{\mathcal{S}\cup {\mathcal{E}}}(t)=e^{-itH_{\mathcal{S}\cup {\mathcal{E%
}}}}\rho _{\mathcal{S}\cup {\mathcal{E}}}(0)e^{itH_{\mathcal{S}\cup {%
\mathcal{E}}}}  \label{dmg}
\end{equation}%
be the density matrix of the composite $\mathcal{S}\cup \mathcal{E}$ (qubits
plus environment), $H_{\mathcal{S}\cup \mathcal{E}}$ be its Hamiltonian and $%
\rho _{\mathcal{S}\cup \mathcal{E}}(0)$ be its initial state.

Following
again a widely used pattern, we will assume that the qubits and the
environment are unentangled initially and that the state of the environment is pure, i.e.,%
\begin{equation}
\rho _{\mathcal{S}\cup \mathcal{E}}(0)=\rho _{\mathcal{S}}(0)\otimes P_{%
\mathcal{E}},\;\ P_{\mathcal{E}}=|\Psi _{\mathcal{E}}\rangle \langle \Psi _{%
\mathcal{E}}|,  \label{icg}
\end{equation}%
where $\rho _{\mathcal{S}}(0)$ is the initial density matrix
of two qubits, a $4\times 4$ positive definite and trace one matrix.

The reduced density matrix of $\mathcal{S}$ (qubits) is then%
\begin{equation}
\rho _{\mathcal{S}}(t)=\mathrm{Tr}_{\mathcal{E}}\rho _{\mathcal{S}\cup {%
\mathcal{E}}}(t),  \label{rdm}
\end{equation}%
where $\mathrm{Tr}_{\mathcal{E}}$ denotes the partial trace with respect to
the degrees of freedom of $\mathcal{E}$.

The linear relation between $\rho _{\mathcal{S}}(t)$ and $\rho _{\mathcal{S}%
}(0)$ given by  (\ref{dmg}) -- (\ref{rdm}) can be written as
\begin{equation}
\rho _{\mathcal{S}}(t)=\Phi (t)\rho _{\mathcal{S}}(0),  \label{super}
\end{equation}%
where the linear superoperator $\Phi (t)$ is known as the quantum channel
superoperator in quantum information theory and is analogous to the influence
(Feynman-Vernon) functional in the theory of open systems.

According to (\ref{dmg}) -- (\ref{rdm}), we obtain a specific model of the
qubit evolution by choosing certain $H_{\mathcal{S}\cup \mathcal{E}}$, $%
\Psi_{\mathcal{E}}$ and $\rho_{\mathcal{S}}$.

\subsection{Hamiltonian}

We will start with the following general form of the Hamiltonian of the
system $\mathcal{S}$ of two qubits embedded into an environment $\mathcal{E}$%
:
\begin{equation}
H_{\mathcal{S}\cup {\mathcal{E}}}=H_{\mathcal{S}}\otimes \mathbf{1}_{%
\mathcal{E}}+\mathbf{1}_{\mathcal{S}}\otimes H_{\mathcal{E}}+H_{\mathcal{SE}%
}.  \label{htot}
\end{equation}%
Here
\begin{equation}
H_{\mathcal{S}}=s_{A}\sigma _{z}^{A}\otimes \mathbf{1}_{\mathcal{S}%
_{B}}+s_{B}\mathbf{1}_{\mathcal{S}_{A}}\otimes \sigma _{z}^{B}  \label{hqub}
\end{equation}%
is the Hamiltonian of two qubits $\mathcal{S}_{a},\;a=A,B$ (spins, 2-level
systems, etc.) written via the Pauli matrices $\sigma _{z}^{A}$ and $\sigma
_{z}^{B}$ and the parameters $s_{A}$ and $s_{B}$, $H_{\mathcal{E}}$ is the
Hamiltonian of the environments and
\begin{equation}
H_{\mathcal{SE}}=Q_{\mathcal{S}}\otimes J_{\mathcal{E}},  \label{hseg}
\end{equation}%
describes the interaction of
the environment and the qubits, where $Q_{\mathcal{S}}$ is a $4\times 4$ hermitian matrix
and $J_{\mathcal{E}}$ is a hermitian matrix acting in the state space of the environment.

For the system $\mathcal{S}=\mathcal{S}_A \cup \mathcal{S}_B$ of two qubits we will choose
\begin{equation}
Q_{\mathcal{S}}=v_{A}\sigma _{x}^{A}\otimes \mathbf{1}_{\mathcal{S}%
_{B}}+v_{B}\mathbf{1}_{\mathcal{S}_{A}}\otimes \sigma _{x}^{B}  \label{qs}
\end{equation}%
with the qubit-environment coupling constants $v_{A}$ and $v_{B}$.

We will indicate now $H_{\mathcal{E}}$ and $J_{\mathcal{E}}$ for our model.
Let $M_{N}$ be a $N\times N$ hermitian matrix (random or not), $%
\{E_{j}^{(N)}\}_{j=1}^{N}$ be its eigenvalues and
\begin{equation}
\nu ^{(N)}(E)=N^{-1}\sum_{j=1}^{N}\delta (E-E_{j}^{(N)})\rightarrow \nu
_{0}(E),\;N\rightarrow \infty  \label{non}
\end{equation}%
be its density of states where $\nu _{0}$ assumed to be continuous and the
limit is understood as the weak limit of measures if $M_{N}$ is not random.
If $M_{N}$ is random, then we assumers that the sequence $\{M_{N}\}_{N}$ is
defined on the same probability space, that the weak convergence holds with
probability 1 in this space and that $\nu _{0}$ is not random, see Section
2.4 of \cite{Pa-Sh:11} for details. For instance, the role of $M_{N}$ can
play matrices studied in Chapter 2 and Sections 7.2, 10.1, 18.3, and 19.2 of
\cite{Pa-Sh:11}.

Furthermore, let $W_{N}$ be a random $N\times N$ hermitian matrix
distributed according to the matrix Gaussian law given by probability
density
\begin{equation}
Z_{N}^{-1}\exp \left\{ -N\mathrm{Tr\;}W_{N}^{2}/2\right\} .  \label{lGOE}
\end{equation}%
where $Z_{N}$ is the normalization constant. In other words, the entries of $%
W_{N}=\{W_{jk}\}_{j,k=1}^{N},\;W_{kj}=W_{jk}^*$ are independent for
$1\leq j\leq k\leq N$ complex Gaussian random variables such that
\begin{equation}
\mathbf{E}\{W_{jk}\}=\mathbf{E}\{W_{jk}^{2}\}=\mathbf{E}%
\{(W_{jk}^*)^{2}\}=0,\; \mathbf{E}\{|W_{jk}|^{2}\}(1+\delta _{jk})/N,  \label{GOE}
\end{equation}%
where $\mathbf{E}\{...\}$ denotes the expectation and the $*$ denotes the complex conjugate.
This is known as the Gaussian Unitary Ensemble (see e.g. \cite%
{Ak-Co:11,Fo:10,Pa-Sh:11}).

We set
\begin{equation}  \label{hehse}
H_{\mathcal{E}}=M_N, \;\; J_{\mathcal{E}}= W_{N}.
\end{equation}
Combining this with (\ref{htot}), (\ref{hqub}) and (\ref{hehse}), we obtain
the Hamiltonian
\begin{equation}
H_{C}=H_{\mathcal{S}}\otimes \mathbf{1}_{\mathcal{E%
}}+\mathbf{1}_{\mathcal{S}}\otimes M_{N}+Q_{\mathcal{S}}\otimes W_{N}^{%
\mathcal{E}},  \label{hc1}
\end{equation}%
of our model of two qubits interacting with a \emph{common} random matrix environment.

We recall also the Hamiltonian $H_{I}$ of the models where each
qubit interacts with its "own" environment and the Hamiltonian $H_{F}$ of the model where one of qubits is free
mentioned in item (i) and (ii) of Introduction.

(i) Hamiltonian $H_I$:
\begin{align}  \label{hi1}
H_{I}&=H_{\mathcal{Q}_A} \otimes \mathbf{1}_{\mathcal{Q}_B} + \mathbf{1}_{\mathcal{Q}_A} \otimes H_{\mathcal{Q}_B}, \; \mathcal{Q}_a=\mathcal{S}_a \cup \mathcal{E}_a, \; a=A,B, \\
H_{\mathcal{Q}_a} &=s_a \sigma_z^a \otimes \mathbf{1} _{\mathcal{E}_a}+\mathbf{1} _{%
\mathcal{S}_a}\otimes M_N^{\mathcal{E}_a} + v_a \sigma_x^a \otimes W_N^{%
\mathcal{E}_a}, \; a=A,B,  \notag
\end{align}
where $M_N^{\mathcal{E}a}, \; a=A,B$ are hermitian matrices satisfying (\ref%
{non}) and $W_{N}^{\mathcal{E}_{a}}, \; a=A,B$ are two hermitian independent
random matrices with the probability distribution (\ref{GOE}). In other
words, every qubit has its own environment and its own interaction with the
environment, hence, the qubits are dynamically \emph{independent}. Here the
entanglement between the qubits for $t> 0$ arises only because they are
initially entangled (see the initial conditions (\ref{1} -- (\ref{3}))
below). The Hamiltonian (\ref{hi1}) can describe two initially entangled and
excited two-level atoms spontaneously emitting into two different cavities,
two sufficiently well separated impurity spins, say, nitrogen vacancy
centers in a diamond microcrystal, etc.

(ii) Hamiltonian $H_F$:
\begin{equation}
H_{F}=H_{\mathcal{S}}\otimes \mathbf{1}_{\mathcal{E}}+\mathbf{1}_{\mathcal{S}%
}\otimes \mathbf{1}_{\mathcal{E}_{A}}\otimes M_{N}^{\mathcal{E}_B}+ \mathbf{1}_{\mathcal{S}%
_{A}} \otimes \mathbf{1}_{\mathcal{E}_{A}} \otimes v\sigma _{x}^{B} \otimes
W_{N}^{\mathcal{E}_{B}}.  \label{hf1}
\end{equation}
i.e., the first qubit is \emph{free} ($H_{\mathcal{S}%
_A \mathcal{E}_A}=0$), but the second qubit is as in (\ref{hf1}). Here also
the qubits do not interact and their quantum correlations for $t> 0$ are due their initial
entanglement (see initial conditions (\ref{1}) -- (\ref{3}) below ). The free qubit is known
as the ancilla or spectator in certain contexts of
quantum information theory, see e.g. \cite{Da-Co:12,El-Si:14,Ho-Co:09,Ri-Co:14}.

These cases were analyzed in detail in our work \cite{Br-Pa:18} and
are used in this work for the comparison of the results pertinent to $H_{I}$ and
$H_{F}$ on one hand and those pertinent to $H_{C}$ on the other hand, since
in the latter case the quantum correlations between the qubits for $t>0$ are not only
due the entangled initial conditions but also due to the interaction, although
indirect, via the environment, between the qubits.

Note that from the point of view of statistical mechanics and condensed
matter theory the Hamiltonians $H_{I}$ and $H_{F}$ of (\ref{hi1}) and (\ref%
{hf1}) seem less interesting than the Hamiltonian $H_{C}$ of (\ref{hc1}),
since $H_{I}$ and $H_{F}$ describe non interacting quantum systems. They
are, however, of considerable interest for quantum information theory, since
the dynamics determined by $H_{I}$ and $H_{F}$ allow for the study of the
emergence of quantum correlations in a "pure kinetic" form, i.e., without
dynamical correlations due to the indirect interaction between the qubits
via the environment as in $H_{C}$ case.

In particular, it seems that the Hamiltonian $H_{I}$ could be a simple model
appropriate for quantum computing, where qubits are independent in typical
solid state devices. Besides, the dynamical independence of qubits can
describe the absence of non-local operations in the quantum information
protocols.

Note also that our Hamiltonians (\ref{hc1}) -- (\ref{hf1}) are the random matrix analogs of widely used
spin-boson Hamiltonians in which the environment Hamiltonian is that of free boson field and the operator
$J_{\mathcal{E}}$ is a linear form in bosonic operators of creation and annihilation,
see \cite{Br-Pe:07,de-Al:17,Le-Co:80,Lo-Co:13,We:08}.

We will discuss new features of the qubits dynamics determined by
Hamiltonian $H_{C}$ in the next sections. Here we note that from the
technical point of view this case is more involved, since, unlike
the Hamiltonians  $H_{I}$ and $H_{F}$, the
channel superoperator for $H_{C}$ is not the tensor product of
the channel operators of
independent qubits but has to be found anew. This is carried out in Section 4.

\subsection{Initial Conditions}

We describe now the initial conditions (\ref{icg}). We will assume that the
pure state $|\Psi _{\mathcal{E}}\rangle $ of environment in (\ref{icg}) is
the eigenstate
\begin{equation}  \label{psik}
|\Psi _{\mathcal{E}}\rangle=\Psi^{(N)}_{k_N}
\end{equation}
of the environment Hamiltonian $H_{\mathcal{E}}=M_{N}$ corresponding to its
eigenvalue $E_{k_N}^{(N)}$ (see (\ref{non}) -- (\ref{hc1})) and that there exist a sequence $\{k_N\}_N$ such that
\begin{equation}
\lim_{N\rightarrow \infty }E_{k_N}^{(N)}=E,\;E\in \mathrm{supp}\;\nu _{0},
\label{ele}
\end{equation}%
see (\ref{non}). Thus, we will denote
\begin{equation}\label{rknn}
\rho_{\mathcal{S}}^{(k_N)}(t)
\end{equation}
the reduced density matrix of two qubits corresponding to the Hamiltonian (\ref{hc1})
and the environment initial condition (\ref{psik}).


As for the initial condition $\rho _{\mathcal{S}}(0)$ for the qubits, we
note that in this paper we obtain the large $N$ limit of the reduced density matrix for any  $\rho _{\mathcal{S}}(0)$.
However, we present below
a rather detailed analysis of the qubit evolution for several initial conditions
that have been considered in a variety of recent papers (see e.g. reviews
\cite{Ao-Co:15,Br-Co:15,Lo-Co:13} and references therein).

We write below $\left\vert
a_{1}a_{2}\right\rangle ,\;a_{1,2}=\pm $ for the vectors $\left\vert
a_{1}\right\rangle \otimes \left\vert a_{2}\right\rangle $ of the standard
product basis of the state space of two qubits where $\left\vert
a\right\rangle ,\;a=\pm $ are the basis vectors of the state space of one
qubit. We also omit the subindex $\mathcal{S}$ in the reduced density
matrices below.

\smallskip
(0) \emph{Condition 0}. The product (hence unentangled) states
\begin{equation}  \label{0}
\rho_0=\rho_A \otimes \rho_B, \; \rho_A=\rho_B=\mathrm{diag}(\alpha_0^2,
1-\alpha_0^2), \; \alpha_0 \in [0,1].
\end{equation}

\smallskip
(i) \emph{Condition 1}. The pure states%
\begin{equation}
\rho _{\Psi _{1}}=\left\vert \Psi _{1}\right\rangle \left\langle \Psi
_{1}\right\vert ,\;\left\vert \Psi _{1}\right\rangle =\alpha _{1}\left\vert
-+\right\rangle +\beta _{1}\left\vert +-\right\rangle ,\;\alpha
_{1}^{2}+|\beta _{1}|^{2}=1  \label{1}
\end{equation}%
known as the Bell-like states and becoming the genuine (maximally entangled)
Bell state if $\alpha _{1}=\beta _{1}=1/\sqrt{2}$.

\smallskip
(2) \emph{Condition 2}. The pure states
\begin{equation}
\rho _{\Psi _{2}}=\left\vert \Psi _{2}s\right\rangle \left\langle \Psi
_{2}\right\vert ,\;\left\vert \Psi _{2}\right\rangle =\alpha _{2}\left\vert
--\right\rangle +\beta _{2}\left\vert ++\right\rangle ,\;\alpha
_{2}^{2}+|\beta _{2}|^{2}=1  \label{2}
\end{equation}%
known also as  Bell-like states and becoming another genuine Bell state for $%
\alpha _{2}=\beta _{2}=1/\sqrt{2}$.

\smallskip
(3) \emph{\ Condition 3(k)}, $k=1,2$. The mixed states%
\begin{equation}
\rho _{W_{k}}=\alpha _{3}\left\vert \Psi _{k}\right\rangle \left\langle \Psi
_{k}\right\vert +((1-\alpha _{3})/4)\mathbf{1}_{4},\;k=1,2,\;-1/3\leq \alpha
_{3}\leq 1.  \label{3}
\end{equation}%
known as the extended Werner states and becoming the genuine Werner state for
$\alpha _{k}=\beta _{k}=1/\sqrt{2},\ k=1,2$. The bound $\alpha _{3}\geq -1/3$
guaranties that $\rho _{W_{k}}$ is positive definite, hence is a state. For $%
\alpha _{3}=1$ $\rho _{W_{k}}$ reduces to $\rho _{\Psi _{k}}$.

The product states (\ref{0}) are always unentangled, the states (\ref{1}) -- (%
\ref{2} are unentangled if
$\alpha _{n}=0,1,\;n=1,2.$
By using the negativity entanglement quantifier (\ref{negx}), it can be
shown that $\rho_{W_{k}}$ of (\ref{3}) is entangled if $1/3<\alpha _{3}\leq 1$ 
and $\alpha_1=\alpha_2=2^{1/2}$.
For other values of $\alpha_1,\alpha_2$ the lower limit $\alpha_3=1/3$ is larger.

In what follows we will call the \textit{model} of the two-qubit evolution
the pair consisting of one of the Hamiltonians (\ref{hc1})  -- (\ref{hf1})
and one of initial conditions (\ref{0}) -- (\ref{3}).
Thus, a particular model is denoted
\begin{equation}
Mm,\;M=C,F,I,\;m=0,1,2,3(k),\ k=1,2,  \label{mod}
\end{equation}%
and for $m=3$ the value of $k=1,2$ from (\ref{3}) has to be indicated.

It is easy to find that in the basis%
\begin{equation}
\left\vert \mathbf{1}\right\rangle =\left\vert ++\right\rangle ,\;\left\vert
\mathbf{2}\right\rangle =\left\vert +-\right\rangle ,\;\left\vert \mathbf{3}%
\right\rangle =\left\vert -+\right\rangle ,\;\left\vert \mathbf{4}%
\right\rangle =\left\vert --\right\rangle \;  \label{2bas1}
\end{equation}%
all the above initial condition have the so-called $X$-form
\begin{equation}
\left(
\begin{array}{cccc}
\rho _{11} & 0 & 0 & \rho _{14} \\
0 & \rho _{22} & \rho _{23} & 0 \\
0 & \rho _{32} & \rho _{33} & 0 \\
\rho _{41} & 0 & 0 & \rho _{44}%
\end{array}%
\right) ,\quad \rho _{32}=\rho _{23}^{\ast },\;\rho _{41}=\rho _{14}^{\ast },
\label{x}
\end{equation}%
which arises in a number of physical situations and is maintained during
widely used dynamics (see \cite{Ao-Co:15,Lo-Co:13,Yu-Eb:09} for reviews). It
is important that the form is also maintained during the dynamics determined
by our random matrix Hamiltonians (\ref{hc1}) -- (\ref{hf1}).
Note that an equivalent block diagonal form
\begin{equation}
\left(
\begin{array}{cccc}
\rho _{11} & \rho _{14} & 0 & 0 \\
\rho _{41} & \rho _{44} & 0 & 0 \\
0 & 0 & \rho _{22} & \rho _{23} \\
0 & 0 & \rho _{32} & \rho _{33}%
\end{array}%
\right) ,\quad \rho _{32}=\rho _{23}^{\ast },\;\rho _{41}=\rho _{14}^{\ast }.
\label{bloc}
\end{equation}%
corresponding to the basis (cf. (\ref{2bas1}))
\begin{equation}
\left\vert \mathbf{1^{\prime }}\right\rangle =\left\vert ++\right\rangle
,\;\left\vert \mathbf{2^{\prime }}\right\rangle =\left\vert --\right\rangle
,\;\left\vert \mathbf{3^{\prime }}\right\rangle =\left\vert +-\right\rangle
,\;\left\vert \mathbf{4^{\prime }}\right\rangle =\left\vert -+\right\rangle
\;  \label{2bas2}
\end{equation}%
is also quite convenient in the analysis of the reduced density matrix of
two qubits. In this case
we will write the $4\times 4$ block matrices (\ref{bloc}), describing two qubits and
their $2\times 2$ diagonal blocks, as follows
\begin{equation}
\left(
\begin{array}{cc}
\rho ^{(+)} & 0 \\
0 & \rho ^{(-)},%
\end{array}%
\right) ,\quad \rho ^{(\eta )}=\{\rho _{\alpha ,\beta }^{(\eta )}\}_{\alpha
,\beta =\pm },\;\eta =\pm .  \label{block}
\end{equation}


\subsection{Quantifiers of Quantum Correlations}

Entanglement, having a short but highly nontrivial mathematical definition
(a state of two quantum objects is entangled if it is not a tensor product
of the states of the objects), is a quite delicate and complex quantum
property admitting a wide variety of physical manifestations and potential
applications. This is also true for general quantum correlations and
motivated the introduction and the active study of a
number of quantitative characteristics (quantifiers, measures, monotones, witnesses)
that are functionals of the corresponding state and
determine the "amount" of its quantum correlations, see reviews \cite%
{Ad-Co:16,Ao-Co:15,Be-Co:17,Br-Co:15,El-Si:14,Ho-Co:09,Lo-Co:13}.
We consider in this paper three
widely used  quantifiers of bipartite states: the negativity,
the concurrence 
and the quantum discord. Since there is a number of reviews and a
considerable amount of original works treating these characteristics, we
give here only their expressions for a two-qubit density matrix
of the $X$ form.

\medskip 
(i) \emph{Negativity} $N[\rho]$ (see reviews \cite%
{Ad-Co:16,Ao-Co:15,El-Si:14,Ho-Co:09})
\begin{align}
&\hspace{1cm}N[\rho]=\max \{0,N_1\}+\max \{0,N_2\},  \label{negx} \\
&N_1=\left( -\rho _{11}-\rho _{44} + \sqrt{(\rho _{11}-\rho
_{44})^{2}+4|\rho _{23}|^{2}}\right),  \notag \\
&N_2=\left( -\rho_{22}-\rho _{33} + \sqrt{(\rho _{22}-\rho _{33})^{2}+4|\rho
_{14}|^{2}}\right).  \notag
\end{align}%
The negativity of a two-qubit state varies from 0 for product states to 1 the maximally
entangled states and is positive
if and only if the state is entangled.

\medskip (ii) \emph{Concurrence} $C[\rho]$ (see reviews \cite%
{Ad-Co:16,Ao-Co:15,El-Si:14,Ho-Co:09,Lo-Co:13},Wo:01)
\begin{align}  \label{conx}
&\hspace{1cm}C[\rho]=2 \ \max\{0,C_1,C_2\}, \\
&C_1=|\rho_{23}| - \sqrt{\rho_{11} \rho_{44}}, \; C_2=|\rho_{14}| - \sqrt{%
\rho_{22} \rho_{33}}.  \notag
\end{align}
The concurrence varies from 0 for separable states to 1 for the maximally entangled
states and is positive if and only if the state is entangled.

The concurrence is one of the most used entanglement quantifier of two-qubit
states, closely related to another entanglement quantifier, known as the
entanglement of formation and applicable in general to multiqubit systems.

Let us mention useful facts on the negativity (\ref{negx}) and the
concurrence (\ref{conx}) of the two-qubit states of $X$-form which can be
easily obtained from (\ref{negx}) and (\ref{conx}).

\smallskip
- $C[\rho ]$ and $N[\rho ]$ are simultaneously positive and
simultaneously vanish, i.e.,%
\begin{equation}
C[\rho ]=0\Longleftrightarrow N[\rho ]=0.  \label{cn0}
\end{equation}

- We have in general
\begin{equation}
C[\rho ]-N[\rho ]\geq 0,  \label{c2n}
\end{equation}%
and the equality%
\begin{equation}
C[\rho ]=N[\rho ]  \label{c2n0}
\end{equation}%
is possible if and only if either $C=C_{1}$ in (\ref{conx}) and $\rho
_{11}=\rho _{44}$ or $C=C_{2}$ in (\ref{conx}) and $\rho _{22}=\rho _{33}$.
In particular, this is the case if the state is pure (see e.g. \cite{El-Si:14,
Wo:01} for the validity of the above relations for other states).

The examples of validity of the above relations are given in \cite{Br-Pa:18} for
the qubit dynamics determined by the Hamiltonians $H_{I}$ of (\ref{hi1}) and
$H_{F}$ of (\ref{hf1}), see Fig. 2b) and 3a) in \cite{Br-Pa:18} and for the Hamiltonian $H_{C}$ of (\ref{hc1}),
see Fig. 1(a) and 2(a) below. Note that in \cite{Br-Pa:18} we use the negativity that is twice less than the negativity
\ref{negx} of this paper.

\medskip (iii) \emph{Quantum discord} $D[\rho ]$ (see reviews \cite{Ad-Co:16,Ao-Co:15,Be-Co:17}).
The quantum discord has a rather involved definition based on
the fact that different quantum analogs of equivalent classical
information quantifiers (e.g. the mutual information) are possible
because measurements perturb a quantum system.
Quantum discord is non-negative in general and is positive for the entangled
states. However, there exist unentangled states having a positive discord,
hence not classical. In other words, the quantum discord "feels" a subtle
difference between product states and classical states and can be viewed as
a measure of total non-classical (quantum) correlations including those that
are not captured by the concurrence and the negativity (2 qubits) and the
entanglement of formation (many qubits).
Unfortunately, we are not aware of a compact formula for the quantum discord
of an arbitrary X-state (\ref{x}) similar to (\ref{negx}) and (\ref{conx})
for the negativity and concurrence. 
However, for the states arising in our models we found a semi-empirical
formula that simplifies considerably the numerical analysis, see \cite{Br-Pa:18}. The formula is used in this paper as well.

\medskip
(iv) \emph{von Neumann
entropy} $S[\rho ]$ (see reviews \cite{Ad-Co:16,Ao-Co:15,Ho-Co:09})
\begin{equation}
S[\rho ]=-\mathrm{Tr}\rho \log_2 \rho,  \label{vne}
\end{equation}%
a quantum analog of the classical Gibbs-Shannon entropy. The von Neumann entropy and its various
modifications play a quite important role in quantum physics ranging from cosmology to biophysics.
In particular, it is a quantifier of the "mixedness" of a quantum state and is also instrumental, together
with certain optimization procedures, in the
definition of various quantum correlation quantifiers, the concurrence and the discord in particular.

Denoting $\{\rho _{\alpha }\}_{\alpha =1}^{4}$ the eigenvalues of the $%
4\times 4$ matrix (\ref{x}), or (\ref{bloc}), we obtain%
\begin{equation}
S[\rho ]=-\sum_{\alpha =1}^{4}\rho _{\alpha }\log_2 \rho _{\alpha },
\label{vnee}
\end{equation}%
where%
\begin{eqnarray}
\rho _{1,4} &=&2^{-1}\left( (\rho _{11}+\rho _{44})\pm \sqrt{(\rho
_{11}-\rho _{44})^{2}+4|\rho _{14}|^{2}}\right), \notag \\
\rho _{2,3} &=&2^{-1}\left( (\rho _{22}+\rho _{33})\pm \sqrt{(\rho
_{22}-\rho _{33})^{2}+4|\rho _{23}|^{2}}\right) \label{vneeg}.
\end{eqnarray}

\section{Results}

\subsection{Analytical Results}

We begin with a convention. We do not indicate explicitly above and below
the dependence on $N$, the number of "degrees of freedom" of the entanglement,
of various objects which include the environment defined via (\ref{non}) -- (\ref{hehse}),
except the cases where it is apparently necessary.

Here is one of the cases.
Since the Hamiltonian (\ref{hc1}) is random because of
(explicitly) random $W_{N}$ and (implicitly) random $M_{N}$, the
corresponding reduced density matrix (\ref{rdm}) is also random. In general,
the complete description of randomly fluctuating objects is given by their
probability distribution. It turns out, however, that in our models the
fluctuations of $\rho _{\mathcal{S}}(t)$ vanish as $N\rightarrow \infty $.
This property is analogous to those known as the representativity of means in
statistical mechanics of macroscopic systems \cite{La-Li}, as the
selfaveraging property in the theory of disordered systems \cite%
{Di-Co:98,LGP} and has been recently discussed in the quantum information
theory \cite{Br-Pa:18,Da-Co:14}.

It is shown in Section 4 (see Result 1) that in the general case of a "$p$%
"-level system, i.e., for the version (\ref{hp}) of $H_{hc1}$ with arbitrary $N$-independent $p\times p$ hermitian $%
H_{\mathcal{S}}$ and $Q_{\mathcal{S}}$,
we have the bound (\ref{sag}). Thus, we can write for the variance of the
entries $(\rho _{\mathcal{S}}(t))_{\alpha \beta },\;\alpha ,\beta
=1,...,4 $ of the reduced density matrix in our case where $p=4$ and $H_{S}$ and $Q_{S}$ are given by (\ref{hqub}) and (\ref{qs}):
\begin{eqnarray}
\mathbf{Var}\{(\rho _{\mathcal{S}}(t))_{\alpha \beta }\}&= &\mathbf{E}%
\{|(\rho _{\mathcal{S}}(t))_{\alpha \beta }|^{2}\}-|\mathbf{E}\{(\rho _{%
\mathcal{S}}(t))_{\alpha \beta }\}|^{2}  \notag \\
&&\leq Ct^{2}/N,\;C=4^{4}(v_{A}+v_{B})^{2}.  \label{sa}
\end{eqnarray}%
Since $N^{-1}$ is the order of magnitude of typical eigenvalue spacings of $%
H_{\mathcal{S}\cup \mathcal{E}}$, we conclude that the order of magnitude of the Heisenberg time for our quantum system (an analog of the Poincar\'{e}
time for classical dynamical systems) is of the order $N$. Thus,
the fluctuations of the reduced density matrix are negligible
if the evolution time of the
system is much less than the Heisenberg time of the system. Note that
analogous condition is well known in non-equilibrium statistical mechanics
as the condition of validity of kinetic regime of macroscopic systems.

The above implies that for large $N$ it suffices to consider the expectation
of the reduced density matrix. The expectation is computed in Section 4 for
a "$p$-level" version (\ref{hp}) of Hamiltonian (\ref{hc1}) in which $H_{S}$ and $Q_{S}$
are arbitrary $N$-independent $p\times p$ hermitian matrices, see Result 2.

Denote
\begin{equation}\label{rknex}
\rho (E,t)=\lim_{N\rightarrow \infty ,\;E_{k_{N}}^{(N)}\rightarrow E}\mathbf{E}\{\rho _{%
\mathcal{S}}^{(k_N)}(t)\}
\end{equation}%
the limit (see (\ref{ele})) of the expectation of the reduced density matrix (\ref{rknn}) corresponding to the Hamiltonian (\ref{hc1}) and the pure state of environment given by (\ref%
{psik}). Then, using Results 2 of Section 4 with $p=4$ and with $H_{S}$ and $%
Q_{S}$ from (\ref{hqub}) and (\ref{qs}), we obtain $\rho (E,t)$ from (\ref%
{rhosg1}) -- (\ref{gz1}).

However, the obtained formulas for $\rho (E,t)$ are not too simple to
analyze effectively both analytically and numerically. To simplify the
formulas, we will first assume that the qubits are identical%
\begin{equation}
s_{A}=s_{B}=s,\;\;v_{A}=v_{B}=v,  \label{sveq}
\end{equation}%
and then pass to the basis (\ref{2bas2}), (see (\ref{bloc}) and (\ref{block}%
)).

In this basis $H_{\mathcal{S}}$ of (\ref{hqub}) is block diagonal while $Q_{%
\mathcal{S}}$ of (\ref{qs}) is block "antidiagonal", i.e.,
\begin{equation*}
H_{\mathcal{S}}=\left(
\begin{array}{cc}
H^{(+)} & 0 \\
0 & H^{(-)},%
\end{array}%
\right) ,\;Q_{\mathcal{S}}=\left(
\begin{array}{cc}
0 & Q \\
Q & 0%
\end{array}%
\right) ,
\end{equation*}%
and
\begin{equation*}
H^{(\eta )}=s(1+\eta 1)\sigma _{z},\;Q=v(1+\sigma _{x}), \; \eta=\pm.
\end{equation*}%
It can be shown that with the above $H_{\mathcal{S}}$ and $Q_{\mathcal{S}}$
the $4\times 4$ matrix $G(E,z)$ in (\ref{gez1}) is block diagonal, i.e., $%
G(E,z)=\{G^{(\eta )}(E,z)\}_{\eta =\pm }$ (see formulas (\ref{Gga}) -- (\ref%
{mfa}) below for its explicit form). This and the block form (\ref{block}) of the
initial conditions $\rho (0)=\{\rho ^{(\eta )}(0)\}_{\eta =\pm }$ in (\ref{0}) -- (\ref{3}) yield the
same form of the $4\times 4$ version of $F_{0}(E,z)=\{F_{0}^{(\eta
)}(E,z)\}_{\eta =\pm }$ in (\ref{fo1}) and then the $4\times 4$ version of (\ref%
{brevG1}) implies the same form of $F(E,z)=\{F^{(\eta )}(E,z)\}_{\eta =\pm }$,
hence of the limiting reduced density matrix $\rho (E,t)=\{\rho ^{(\eta
)}(E,t)\}_{\eta =\pm }$ in (\ref{rhosg1}).

To write down the obtained block form of our basic equations (\ref{rhosg1}) -- (\ref{gz1}) for $p=4$, the two
qubits case of (\ref{hp}), it is convenient to introduce for any $2\times 2$
matrix $A=\{A_{\alpha ,\beta }\}_{\alpha ,\beta =\pm }$ the number
\begin{equation*}
\mathcal{T}(A)=\sum_{\alpha ,\beta =\pm }A_{\alpha ,\beta }=\mathrm{Tr}%
A(1+\sigma _{x}).
\end{equation*}%
We have then after a certain amount of linear algebra
\begin{equation}
\rho ^{(\eta )}(E,t)=-\frac{1}{(2\pi i)^{2}}\int_{-\infty -i \varepsilon}^{\infty -i \varepsilon}dz_{1}%
\int_{-\infty +i \varepsilon}^{\infty +i \varepsilon}dz_{2}e^{i(z_{1}-z_{2})t}F^{(\eta )}(E,z_{1},z_{2}),\;\eta =\pm ,
\label{rho2}
\end{equation}%
with
\begin{align}
& F^{(\eta )}(E,z_{1},z_{2})=F_{0}^{(\eta )}(E,z_{1},z_{2})  \notag \\
& \hspace{0.5cm}+v^{2}G^{(\eta )}(z_{1},z_{2})\frac{\mathcal{F}_{0}^{(-\eta
)}(E,z_{1},z_{2})+v^{2}\mathcal{F}_{0}^{(\eta )}(E,z_{1},z_{2})\mathcal{G}%
^{(-\eta )}(z_{1},z_{2})}{1-v^{4}\mathcal{G}^{(+)}(z_{1},z_{2})\mathcal{G}%
^{(-)}(z_{1},z_{2})},\;\eta =\pm ,  \label{fg}
\end{align}%
where
\begin{eqnarray}
\mathcal{F}_{0}^{(\gamma )}(E,z_{1},z_{2}) &=&\mathcal{T}(F_{0}^{(\eta
)}(E,z_{1},z_{2})),\;\mathcal{F}^{(\eta )}(E,z_{1},z_{2})=\mathcal{T}%
(F^{(\eta )}(E,z_{1},z_{2})  \notag \\
\mathcal{G}^{(\eta )}(z_{1},z_{2}) &=&\mathcal{T}(G^{(\eta
)}(z_{1},z_{2})),\quad \mathcal{G}^{(\eta )}(E,z)=\mathcal{T}(G^{(\eta
)}(E,z)),\;\eta =\pm .  \label{tcs}
\end{eqnarray}
and
\begin{align}
& F_{0}^{(\eta )}(E;z_{1},z_{2})=G^{(\eta )}(E,z_{2})\rho ^{(\eta
)}(0)G^{(\eta )}(E,z_{1}),  \notag \\
& G^{(\eta )}(z_{1},z_{2})=v^2\int G^{(\eta )}(E,z_{2})(1+\sigma _{x})G^{(\eta
)}(E,z_{1})\nu _{0}(E)dE,\;\eta =\pm ,  \label{f0g}
\end{align}%
in which
\begin{align}
& G^{(\eta )}(E,z)=\frac{E-z\sigma _{x}-s(1+\eta 1)\sigma _{z}-Z^{(-\eta
)}(z)(1-\sigma _{x})}{E^{2}-z^{2}-4s^{2}-2(E-z)Z^{(-\eta )}(z)}  \label{Gga}
\\
& Z^{(\eta )}(z)=z+v^{2}\mathcal{G}^{(\eta )}(z),\;\mathcal{G}^{(\eta
)}(z)=\int \mathcal{G}^{(\eta )}(E,z)\nu _{0}(E)dE  \label{gz}
\end{align}%
and the pair $\{\mathcal{G}^{(\eta )}(z)\}_{\eta =\pm }$ solves uniquely the
equation%
\begin{equation}
\mathcal{G}^{(\eta )}(z)=\int \frac{2(E-z)\nu _{0}(E)dE}{%
E^{2}-z^{2}-4s^{2}-2
(E-z)Z^{(-\eta )}(z)},\;\eta =\pm.  \label{mfa}
\end{equation}%
in the class of $2\times 2$ matrix functions analytic for $\Im z\neq 0$ and
satisfying (\ref{neva}) for $p=2$.

Note that (\ref{mfa}) can be viewed as an analog of selfconsistent equations
of the mean field approximation in statistical mechanics (recall the
Curie-Weiss and van der Waals equations). In fact, it is widely believed
that random matrices of large size provide a kind of mean field models for
the one body disordered quantum systems. Correspondingly, random matrix
theory deals with a number of selfconsistent equations, see e.g. \cite{Pa-Sh:11}.

Given the solution of (\ref{mfa}), we obtain $G(E,z)$ from (\ref{Gga}) and
then the integrand in (\ref{rho2}) via (\ref{fg}) -- (\ref{tcs}). Next, we
have to compute the contour integrals in (\ref{rho2}) and to get explicit
formulas for the reduced density matrix. The integrals are determined by the
zeros of the denominator of (\ref{fg}) in $z_{1}\in \mathbb{C}_{+}$ and $%
z_{2}\in \mathbb{C}_{-}$. The corresponding analysis proved to be a quite
non-trivial problem even in the single qubit ($p=1$) case considered in \cite%
{Le-Pa:03}. In that paper we were able to carry out the analysis and to
compute the integrals by using an analog of the so-called Bogolyubov -
van Hove regime where%
\begin{equation}
t\rightarrow \infty ,v\rightarrow 0,\;v^{2}t\rightarrow \tau \in \lbrack
0,\infty ),  \label{bvh}
\end{equation}%
where $\tau $ is known as the slow or coarse-grained time.

The regime is known since the 1930's in the theory of finite dimensional
dynamical systems \cite{Bo-Mi:62} as an efficient modification of the small
nonlinearity perturbation theory valid on the $O(v^{-2})$-time intervals in
contrast to the standard perturbation theory, valid on the $O(v^{-1})$-time
intervals. It was then used by Bogolyubov in the 1940th \cite{Bo:45} to
obtain the Markovian description (via the Ornstein-Uhleneck Markov process) of
the dynamics of a classical oscillator coupled linearly to a macroscopic
environment of classical oscillators and by van Hove in the 1950th \cite{vH:55}
to obtain the kinetic description (via various master equations) of
macroscopic quantum systems. Since then the regime is a basic ingredient
to obtain the Markovian description known also as the Born-Markov approximation
in the theory of open systems and nonequilibrium statistical mechanics \cite%
{Br-Pe:07,Da:76,de-Al:17,Sp:80,We:08} resulting, in particular, in the so called
quantum Brownian motion (Lindblad dynamics). For the applicability and
quantification of the Markov approximation in quantum dynamics of qubits see
\cite{Ad-Co:13,Br-Co:15,Da-Co:12,Ri-Co:14}. In general, the Markovian
description is applicable on the time intervals lying between the
relaxation time of the environment
correlations and the available time of the system's evolution, the former is
assumed to be much shorter than the latter., see e.g. \cite{VK:07}.
The Markov approximation has been successfully used in quantum optics.
On the other hand, it follows from numerous recent works that non-Markovian
effects are of great importance in a wide variety of quantum contexts
ranging from quantum thermodynamics to communication protocols.
As for quantum information theory, it was found that the Markovian regime
leads to the monotone and exponentially vanishing at a finite moment
concurrence and negativity (see e.g. Fig 1a) below)
whereas the non-Markovian regime allows for the revivals of these
entanglement quantifiers thereby predicting a larger and a longer
living entanglement mediated by the backflow of the information from the environment
to the system (see Fig. 2 -- Fig. 4a) and Fig. 5b) below).

The mostly used so far models of non-Markovian dynamics are based on
particular solutions and various approximations of the two-qubit
version of the so-called spin-boson model \cite{Ad-Co:13,Ao-Co:15,de-Al:17,Le-Co:80,Lo-Co:13}.
It was shown in \cite%
{Br-Pa:18,Le-Pa:03} that for the one qubit model with the random matrix
environment the dynamics is not Markovian in general even in the regime (\ref%
{bvh}). For our model of the two qubit dynamics in the common random matrix
environment the formal proof is given below, after formula (\ref{aar}).

We present now the reduced density matrix $\rho (E,\tau )$ of our model
in the regime (\ref{bvh}). The corresponding calculations are just a
somewhat more technically involved version of those in \cite{Le-Pa:03},
Section 5, since the algebraic structure of the $2\times 2$ block formulas (%
\ref{rho2}) -- (\ref{mfa}) is quite similar to that of the scalar ($1\times
1 $) block formulas in \cite{Le-Pa:03}, Section 4. It is necessary to change variables $(z_1,z_2)$ to $z=z_2, \; \zeta=(z_1-z_2)v^{-2}$ in (\ref{rho2}) -- (\ref{tcs}) and then find their limiting form in the regime (\ref{bvh}).

We denote
\begin{align}
& \nu _{\alpha }=\nu _{0}(E+2\alpha s),\;\alpha =0,\pm , \; \Gamma
_{\alpha }=2\pi \nu _{\alpha }, \;  \Gamma
_{2\alpha }=2\pi \nu (E+ 4\alpha s), \;\alpha =\pm, \notag \\
& \widetilde{\Gamma }_{\alpha }=\Gamma _{0}+\Gamma _{\alpha }+\Gamma
_{2\alpha },\quad \Gamma =\sum_{\alpha =\pm }\Gamma _{\alpha },\quad
\widetilde{\Gamma }=\sum_{\alpha =0,\pm }\Gamma _{\alpha },  \label{nuga}
\end{align}%
In this notation we have in the interaction representation
\begin{align}
& \rho ^{(+)}(E,\tau )=\sum_{\alpha =\pm }p_{\alpha }\Big\{(q_{\alpha
}+e^{-2\Gamma _{\alpha }\tau })\left( \rho ^{(+)}(0)\right) _{\alpha \alpha
}+\frac{\Gamma _{-2\alpha }}{\Gamma_0}q_{-\alpha }\left( \rho ^{(+)}(0)\right) _{-\alpha
,-\alpha }  \notag \\
&\hspace{1cm} +\frac{\Gamma _{-\alpha }}{\widetilde{\Gamma }}(1-e^{-2%
\widetilde{\Gamma }\tau })A_{1}\Big\}+2\Re \sigma_{+}e^{i\alpha \Psi _{+}\tau }e^{-\Gamma \tau }\left( \rho ^{(+)}(0)\right)
_{+,-},  \label{rdvh1}
\end{align}
\begin{align}
& \rho ^{(-)}(E,\tau )=\pi _{+}\left[ \sum_{\alpha =\pm }\frac{\Gamma
_{\alpha }}{\widetilde{\Gamma }_{\alpha }}(1-e^{-2\widetilde{\Gamma }%
_{\alpha }\tau })\left( \rho ^{(+)}(0)\right) _{\alpha \alpha }+\Big(\frac{\Gamma
_{0}}{\widetilde{\Gamma}}+\frac{\Gamma _{+}+\Gamma _{-}}{\widetilde{\Gamma }%
}e^{-2\widetilde{\Gamma}\tau }\Big)A_{1}\right] \notag \\
&\hspace{1cm} +\pi _{-}A_{2}+\Re (\sigma _{z}+i\sigma_{y})e^{i\Psi _{-}\tau }e^{-\Gamma \tau }A_{3},
\label{rdvh2}
\end{align}%
where for any $2\times 2$ matrix $A$ we write $\Re A=(A+A^{+})/2$  (cf. (\ref{ima})) and denote
\begin{eqnarray}
p_{\alpha } &=&\frac{1+\alpha \sigma _{z}}{2},\quad \pi _{\alpha }=\frac{%
1+\alpha \sigma _{x}}{2},\; \notag\\
q_{\alpha } &=&\frac{\Gamma _{0}}{\widetilde{\Gamma }_{\alpha }}+\frac{%
\Gamma _{\alpha }\Gamma _{0}}{\widetilde{\Gamma }_{\alpha }(\Gamma
_{0}+\Gamma _{2\alpha })}e^{-2\widetilde{\Gamma }_{\alpha }\tau }-\frac{%
\Gamma _{0}}{\Gamma _{0}+\Gamma _{2\alpha }}e^{-2\Gamma _{\alpha }\tau }.
\label{pgas}
\end{eqnarray}
\begin{equation}
\Psi _{+}=-8\ s\;v.p.\int \frac{\nu _{0}(E^{\prime })}{(E^{\prime
}-E)^{2}-4s^{2}}dE^{\prime },\quad \Psi _{-}=4\;v.p.\int \frac{\nu
_{0}(E^{\prime })(E^{\prime }-E)}{(E^{\prime }-E)^{2}-4s^{2}}dE^{\prime },
\label{psys}
\end{equation}
where $v.p.$ denotes the integral in the Cauchy sense at points where the denominator of the integrand is zero,
\begin{align}
&A_{1}=\frac{1}{2}\sum_{\alpha ,\alpha ^{\prime }=\pm }\left( \rho
^{(-)}(0)\right) _{\alpha \alpha ^{\prime }},\;A_{2}=\frac{1}{2}\sum_{\alpha
,\alpha ^{\prime }=\pm }\alpha \alpha ^{\prime }\left( \rho ^{(-)}(0)\right)
_{\alpha \alpha ^{\prime }},\notag
\\& A_{3}=\frac{1}{2}\sum_{\alpha ,\alpha ^{\prime
}=\pm }\alpha \left( \rho ^{(-)}(0)\right) _{\alpha \alpha ^{\prime }}.
\label{An}
\end{align}%
In particular, we have for the large time limit of the reduced density
matrix
\begin{align}
& \rho _{\infty }^{(+)}=\sum_{\alpha =\pm }\ \ p_{\alpha }\left\{ \frac{%
\Gamma _{0}}{\widetilde{\Gamma }_{\alpha }}\cdot \left( \rho
^{(+)}(0)\right) _{\alpha \alpha }+\frac{\Gamma _{-2\alpha }}{\widetilde{%
\Gamma }_{\alpha }}\cdot \left( \rho ^{(+)}(0)\right) _{-\alpha ,-\alpha }+%
\frac{\Gamma _{-\alpha }}{\widetilde{\Gamma }}A_{1}\right\}  \notag \\
& \rho _{\infty }^{(-)}=\pi _{+}\left( A_{1}\frac{\Gamma _{0}}{\widetilde{%
\Gamma }}+\sum_{\alpha =\pm }\frac{\Gamma _{\alpha }}{\widetilde{\Gamma }%
_{\alpha }}\left( \rho ^{(+)}(0)\right) _{\alpha \alpha }\right) +\pi
_{-}A_{2}.  \label{rinf}
\end{align}
Note that the dependence on the initial conditions of the infinite time
limit of the reduced density matrix (\ref{rinf}) is not typical
for Markovian dynamics. Moreover, we will give now a formal proof that the
dynamics given by (\ref{nuga}) -- (\ref{psys}) is not Markovian generically.

To this end it is convenient to pass from the entries $\rho _{\alpha \beta
}^{(-)},\;\alpha ,\beta =\pm $ of the second block in (\ref{rdvh1}) -- (\ref%
{rdvh2}) to their linear combinations $A_{k},\;k=1,2,3$ given by (\ref{An}).
We obtain
\begin{equation}
\left(
\begin{array}{c}
\rho _{11}(\tau ) \\
A_{1}(\tau ) \\
\rho _{44}(\tau )%
\end{array}%
\right) =\left(
\begin{array}{ccc}
q_{+}+e^{-2\Gamma _{+}\tau },\quad \frac{\Gamma _{-}}{\widetilde{\Gamma }}%
(1-e^{-2\widetilde{\Gamma }\tau }),\quad \frac{\Gamma _{-2}}{\Gamma _{0}}%
q_{-} &  &  \\
\frac{\Gamma _{+}}{\widetilde{\Gamma }_{+}}(1-e^{-2\widetilde{\Gamma }%
_{+}\tau }),\frac{\Gamma _{0}}{\widetilde{\Gamma }}+\frac{\Gamma }{%
\widetilde{\Gamma }}e^{-2\widetilde{\Gamma }\tau },\frac{\Gamma _{-}}{%
\widetilde{\Gamma }_{-}}(1-e^{-2\widetilde{\Gamma }_{-}\tau }) &  &  \\
\frac{\Gamma _{+2}}{\Gamma _{0}}q_{+},\quad \frac{\Gamma _{+}}{\widetilde{%
\Gamma }}(1-e^{-2\widetilde{\Gamma }\tau }),\quad q_{-}+e^{-2\Gamma _{-}\tau
} &  &
\end{array}%
\right) \left(
\begin{array}{c}
\rho _{11}(0) \\
A_{1}(0) \\
\rho _{44}(0)%
\end{array}%
\right) ,  \label{rar}
\end{equation}%
\begin{equation}
\hspace{-0.34cm}A_{2}(\tau )=A_{2}(0),\quad A_{3}(\tau )=e^{-\Gamma \tau
}e^{i\Psi _{2}\tau }A_{3}(0)\quad \rho _{14}(\tau )=e^{-\Gamma \tau
}e^{i\Psi _{1}\tau }\rho _{14}(0).  \label{aar}
\end{equation}
It follows from (\ref{rar}) -- (\ref{aar}) that the dynamics of $(\rho
_{11},A_{1},\rho _{44})$ given by (\ref{rar}) is independent of that of $(A_{2},A_{3},\rho _{14})$ given by (\ref{aar}). Hence,
the corresponding channel operator has a block form with
three $1\times 1$ blocks for $A_{2},A_{3}$ and $\rho _{14})$, each evolving independently, and the $%
3\times 3$ block for $(\rho _{11},A_{1}),\rho _{44})$.

Recall that the Markov evolution of the reduced density matrix corresponding
to a time-independent Hamiltonian is described by the exponential channel
superoperator of (\ref{super}):
\begin{equation}
\Phi (\tau )=e^{-\tau \mathcal{L}},  \label{Marex}
\end{equation}%
see, however, \cite{Br-Co:15,Mi-Co:19,Ri-Co:14} for discussions of quantum Markovianity.

According to (\ref{aar}), the three $1\times 1$ blocks are exponential in $\tau $, except $A_2$
where the evolution is absent because of the special symmetry
of a general Hamiltonian of two qubits with a common environment, see e.g. \cite{Lo-Co:13,Ma-Co:09}. Thus,
the dynamics of $(A_{2},A_{3},\rho _{14})$ satisfies (\ref{Marex}) and we can confine ourselves
to the $3\times 3$  block given by(\ref{rar}),
i.e., to the restriction of the dynamics to the subspace of $(\rho _{11},\rho
_{44},A_{1}) $. Denote $\Phi _{3}$ the restriction of $\Phi $ to this
subspace and assume that $\Phi _{3}$ is exponential, hence,
\begin{equation}
\Phi _{3}(\tau +\tau _{1})=\Phi _{3}(\tau )\Phi _{3}(\tau _{1})
\label{divis}
\end{equation}%
for any $\tau ,\tau _{1}\geq 0$. Then, carrying out the limits $\tau ,\tau
_{1}\rightarrow \infty $, we obtain $\Phi _{3}(\infty )=\Phi _{3}^{2}(\infty
)$. If $\Phi _{3}(\infty )$ is invertible, it is the unity, i.e., the
dynamics is trivial. Hence, a non trivial Markovian dynamics corresponds
to a non invertible $\Phi _{3}(\infty )$ with $\det \Phi _{3}(\infty )=0$.
This is a condition on the density of states $\nu _{0}$ of the environment,
a functional parameter of our model. We conclude that the Markovianity of $%
\Phi _{3}$, hence of our model (\ref{rdvh1}) -- ( \ref{An}), is not generic.
In other words, (\ref{rar}) cannot be obtained in general as a solution of
a system of three ordinary differential equations.

A simple case of the Markovianity of $\Phi _{3}$ in (\ref{rar}) with $\det
\Phi _{3}(\infty )=0$ corresponds to the "locally flat" density of states $\nu_0$ of (\ref{non}),
where  $\nu_0 (E)=\nu_0 (E \pm 2s)=\nu_0(E \pm 4s)$, i.e., see (\ref{nuga})
\begin{equation}\label{fdos}
\Gamma _{0}=\Gamma _{\alpha }=\Gamma _{2\alpha },\;\alpha =\pm.
\end{equation}
It follows
from (\ref{rar}) that in this case $\Phi _{3}(\tau )=e^{-\mathcal{L}_{3}\tau
}$, where $\mathcal{L}_{3}$ is the $3\times 3$ hermitian matrix with
eigenvalues $0,2\Gamma _{0},6\Gamma _{0}$ and eigenvectors $%
e_{1}=3^{-1/2}(1,1,1)$, \ $e_{2}=6^{-1/2}(1,-2,1)$,$\;e_{3}=2^{-1/2}(1,0,-1)$
which is the infinitesimal operator of the three states Markov process \cite{VK:07}. Correspondingly,
the triple ($\rho_{11}, A_1, \rho_{44}$) converges as  $\tau \to \infty$ to the unique stationary
state $e_1$. Moreover, the whole reduced density matrix of two qubits have in this case the unique
stationary maximally mixed state $4^{-1}(1,1,1,1)$.

This has to be compared to the one qubit random matrix model considered in \cite{Br-Pa:18,Le-Pa:03}. There the dynamics of the diagonal entries and the off-diagonal entry of the $2 \times 2$ reduced density matrix are independent in the regime (\ref{bvh}). The off-diagonal entry decays exponentially as $\tau \to \infty$ (cf. (\ref{aar})). The entries of the channel superoperator $\Phi_2(\tau)$ for the diagonal entries are parametrized by  $\nu_{\alpha}, \; \alpha=0,\pm$ (cf. (\ref{nuga}) and (\ref{rar})). The condition $\det \Phi_2(\infty)=0$ is equivalent to $\nu_{+}=\nu_{-}$ while the Markovian dynamics is the case if and only if
\begin{equation}\label{mark1}
\nu_{+}=\nu_{-}=\nu_0,
\end{equation}
which is a natural analog of (\ref{ldos}). The diagonal entries converge exponentially fast to
the unique and independent on the initial conditions stationary state $e_1=2^{-1/2}(1,1)$.

\subsection{Numerical Results}

We present now our results on the numerical analysis of the time
evolution in the regime (\ref{bvh}) of the negativity, the concurrence, the quantum discord and
the entropy for the random matrix models given by
initial conditions (\ref{0}) -- (\ref{3}) and the Hamiltonian (\ref{hc1}) of
two identical qubits both interacting with the same environment and compare
them with analogous results for the Hamiltonian (\ref{hi1}) of two identical
qubits each interacting with its own environment and Hamiltonian (\ref{hf1})
for two identical qubits with only one of them interacting with an
environment.

The results are based on formulas (\ref{nuga}) -- (\ref{rinf}) or (\ref{nuga}) and (\ref{rar}) -- (\ref{aar})
and the Lorenzian density of states
\begin{equation}\label{ldos}
\nu_0 (E)=\frac{\gamma}{\pi (E^2 + \gamma^2)}
\end{equation}

It will also be convenient to use the energy units where the qubit amplitude $s$ of (\ref{sveq}) is set to 1.

Let us recall first that in view of bound (\ref{sa}), providing the
selfaveraging property (typicality) of the reduced density matrices in
question, all the quantifiers are non random in the large $N$ limit.
Note also that in the regime (\ref{bvh}) the r.h.s. of (\ref{sa}) with (\ref{sveq}) is
$O(t \tau / N)$ and the fluctuations of the reduced density matrix, hence, the quantifiers, are negligible if $\tau \ll t \ll N)$.

\begin{figure}[h]
\begin{subfigure}[b]{.49\textwidth}
  \centering
  \includegraphics[width=\linewidth]{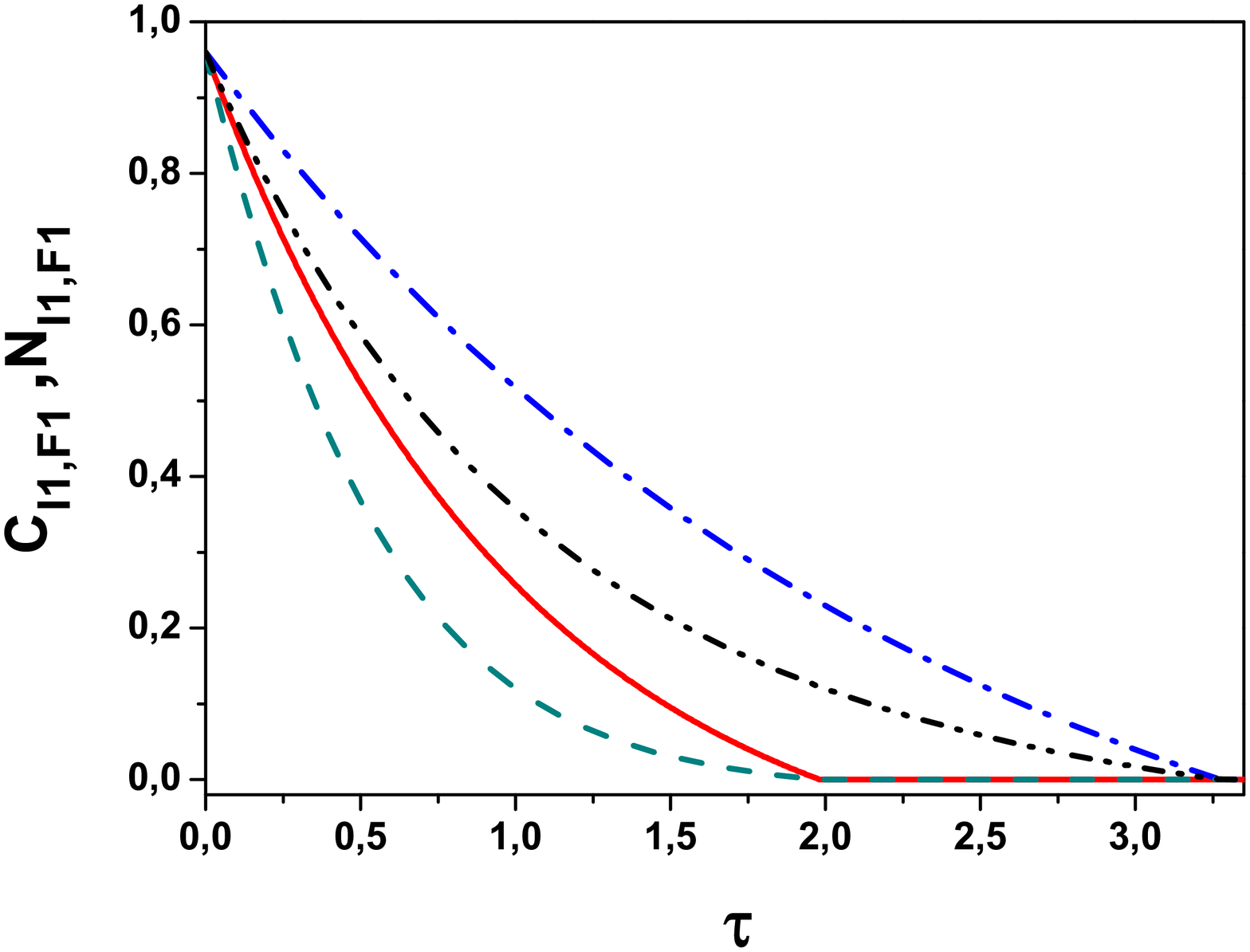}
 \caption{}
\end{subfigure}
\begin{subfigure}[b]{.49\textwidth}
  \centering
  \includegraphics[width=\linewidth]
  {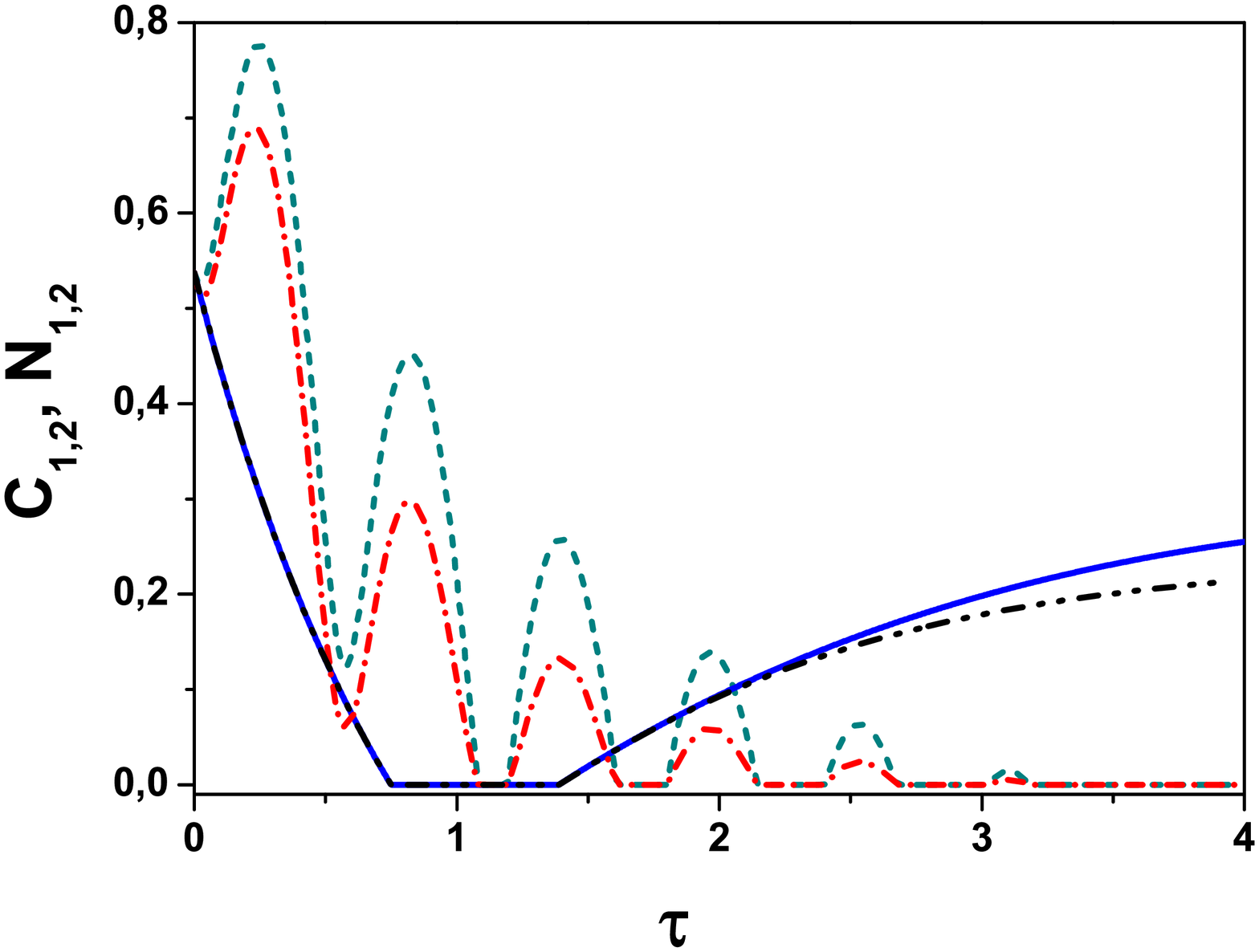}
 \caption{}
\end{subfigure}
\caption{ \small (a)
 Concurrence and negativity  for models $F1$ and $I1$
 (Hamiltonians (\ref{hf1}) and pure Bell-like initial
states (\ref{1}), $\alpha_1=0.6$). The solid red line is $C_{I1}$,
the blue dash dot line is $C_{F1}$, the green dash line is $N_{I1}$ and
the black dash dot dot line is $N_{F1}$. Observe the ESD phenomenon
and that  $C_{M1}=N_{M1}=0, \; \tau=\tau_{ESD}, \; C_{M1}>N_{M1}, \; \tau< \tau_{ESD}, \; M=I,F$, 
cf. (\ref{cn0}) -- (\ref{c2n}).
(b)   Concurrence and negativity for the models $C1$ and $C2$ (Hamiltonian (\ref{hc1}) and Bell-like
pure initial states (\ref{1}) and (\ref{2}) with $\alpha:=\alpha_1=\alpha_2=0.96$) and the
same model parameters ($\Gamma$-s in (\ref{nuga})).
The green short dash line is
$C_{C1}$, the red short dash dot line is $N_{C1}$. Multiple
alternating ESD and ESB with positive minimum $\tau_{ESD}$ and finite maximum $\tau_{ESD}$.
The blue solid
line  is $C_{C2}$, the black dash dot dot line is $N_{C2}$ with $0 <
\tau_{ESD} < \tau_{ESB} < \infty$
and $C_{C2}(\infty)=0.266$  > $N_{C2}(\infty)=0.227$.
}
\end{figure}

Fig. 1(a). The figure is taken from \cite{Br-Pa:18}. It describes the
evolution of the concurrence and the negativity corresponding to
Hamiltonians (\ref{hi1}) -- (\ref{hf1}) and the initial condition
(\ref{1}) and is given here for the comparison.
It shows a simple case of the Entanglement Sudden Death (ESD)
phenomenon \cite{Lo-Co:13,Yu-Eb:04,Zy-Co:01}: the monotone $C > N$ curves for $0<\tau < \tau_{ESD}$,
simultaneous ESD at $\tau=\tau_{ESD}< \infty$ and no the Entanglement
Sudden Birth (ESB) phenomenon \cite{Lo-Co:13,Yu-Eb:09,Zy-Co:01} for larger $\tau$'s (cf. (\ref{c2n}) and
(\ref{c2n0})).  This is a
manifestation of the absence of the inverse flow of information
from the environment to qubits pertinent to the  Markovian models and
preventing the ESB, although our models $I1$ and $F1$ are not Markovian
in general, see (\ref{mark1}) and \cite{Br-Pa:18}. It is worth mentioning that the results shown on Fig. 1(a) are in good
qualitative agreement with certain all-optical experiments, see, e.g. \cite{Ao-Co:15}, Fig. 21.

\begin{figure}[t]
\begin{subfigure}[b]{.49\textwidth}
  \centering
  \includegraphics[width=\linewidth]{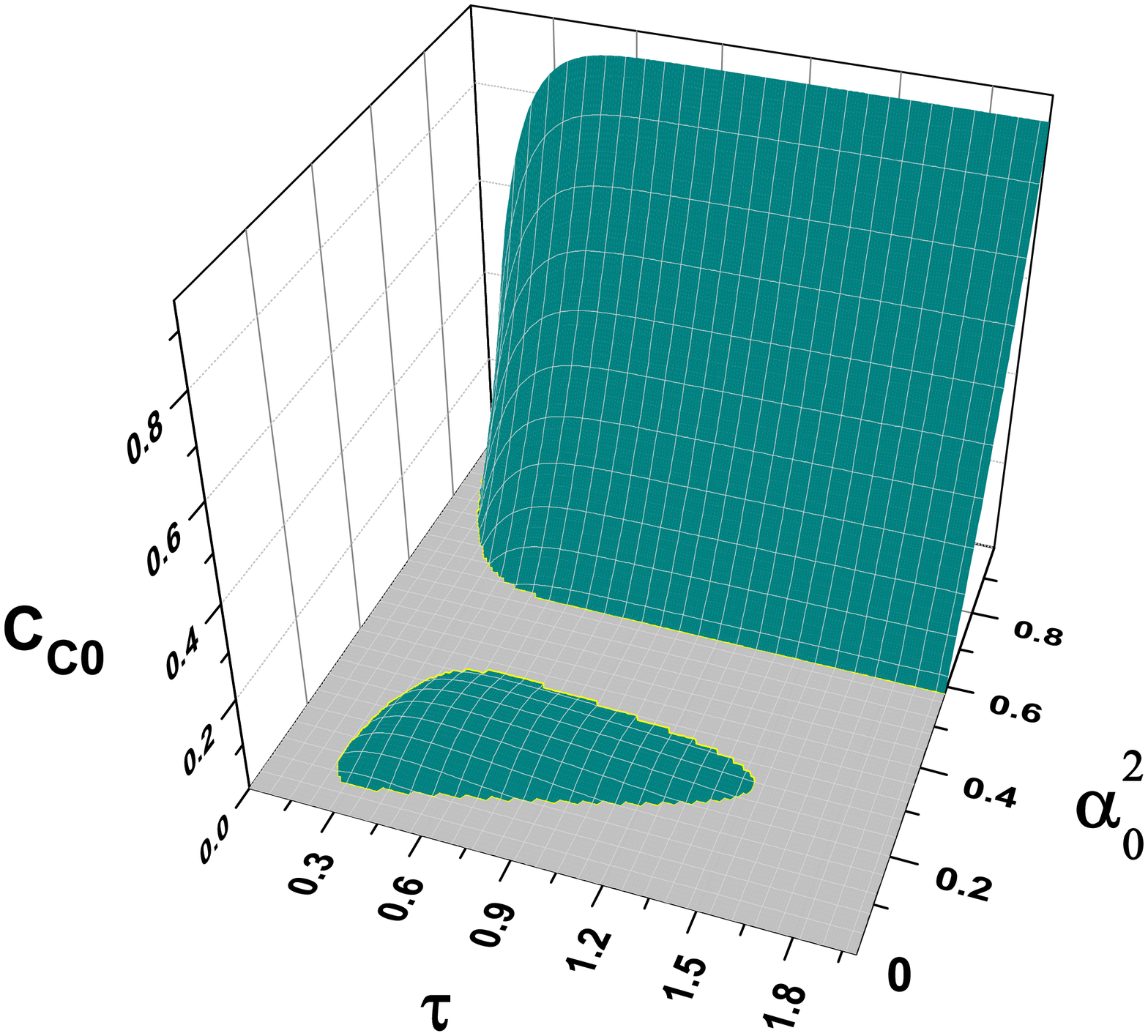}
 \caption{}
\end{subfigure}
\begin{subfigure}[b]{.49\textwidth}
  \centering
  \includegraphics[width=\linewidth]{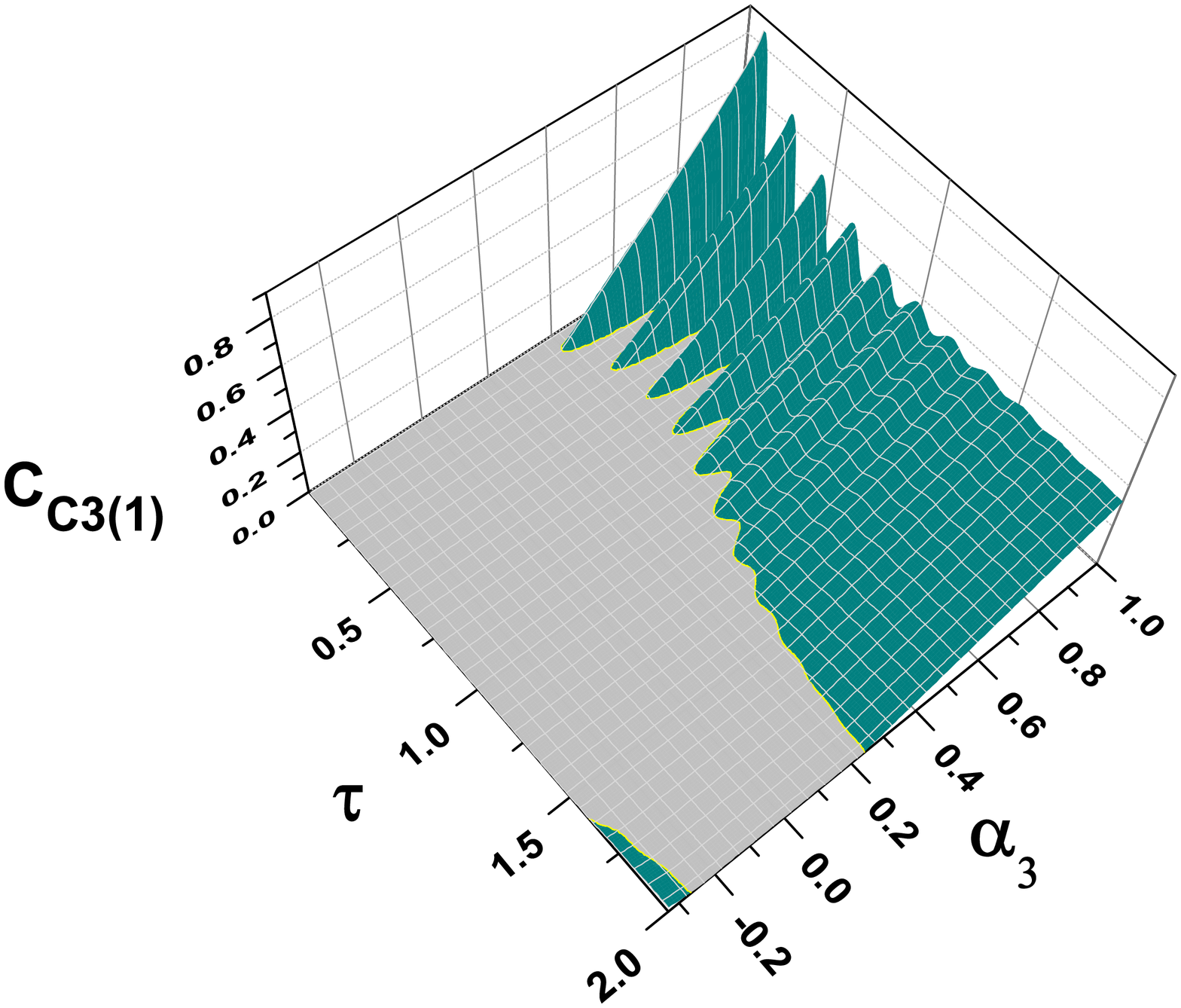}
 \caption{}
\end{subfigure}
\caption{ \small (a)
 Concurrence for the model $C0$ (the product, hence, unentangled initial states for all $\alpha_0 \in [0,1]$).
$\alpha_0$ close to 1 : fast growth for $\tau
>\min \tau_{ESB} > 0$, slow decay for large $\tau$ with zero or non-zero (trapping) at infinity.
Small $\alpha_0$: an "island" of non-zero entanglement with $0 < \tau_{ESB} <
\tau_{ESD} < \infty$.
(b)
Concurrence for the model $C_{3(1)}$ (unentangled for $|\alpha_3| \leq 1/3$ initial states (\ref{3}) and $\alpha_1=0.1$).
Small $\alpha_3$: a finite "island" of non-zero entanglement.
Intermediate $\alpha_3$: no entanglement. Close to 1 $\alpha_3$: $0 <
\tau_{ESD}^{(1)}< \tau_{ESB}^{(2)} < \tau_{ESD}^{(2)}$ finite or infinite.
}
\end{figure}

\begin{figure}[ht]
\begin{subfigure}[b]{.49\textwidth}
  \centering
  \includegraphics[width=\linewidth]{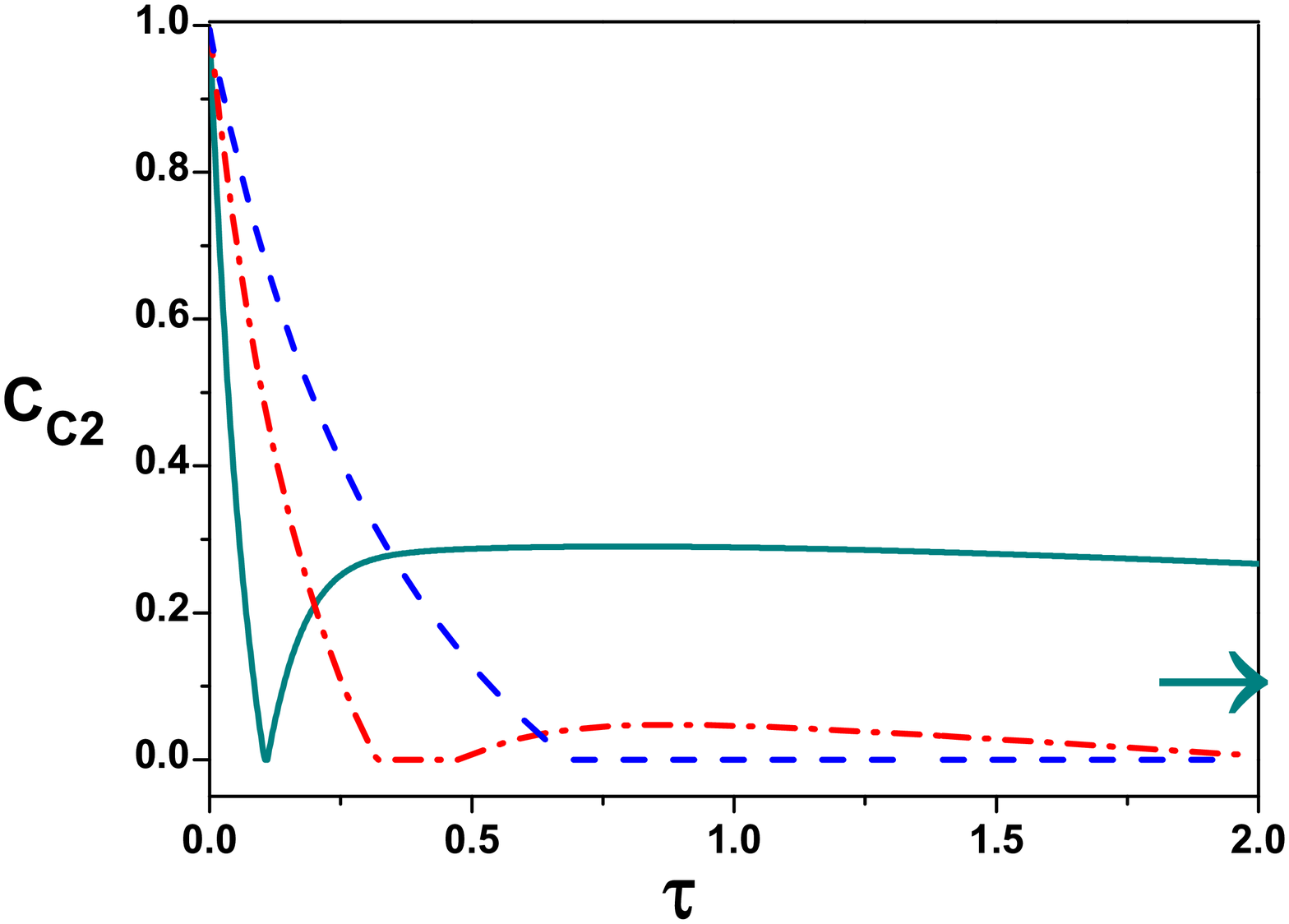}
 \caption{}
\end{subfigure}
\begin{subfigure}[b]{.49\textwidth}
  \centering
  \includegraphics[width=\linewidth]{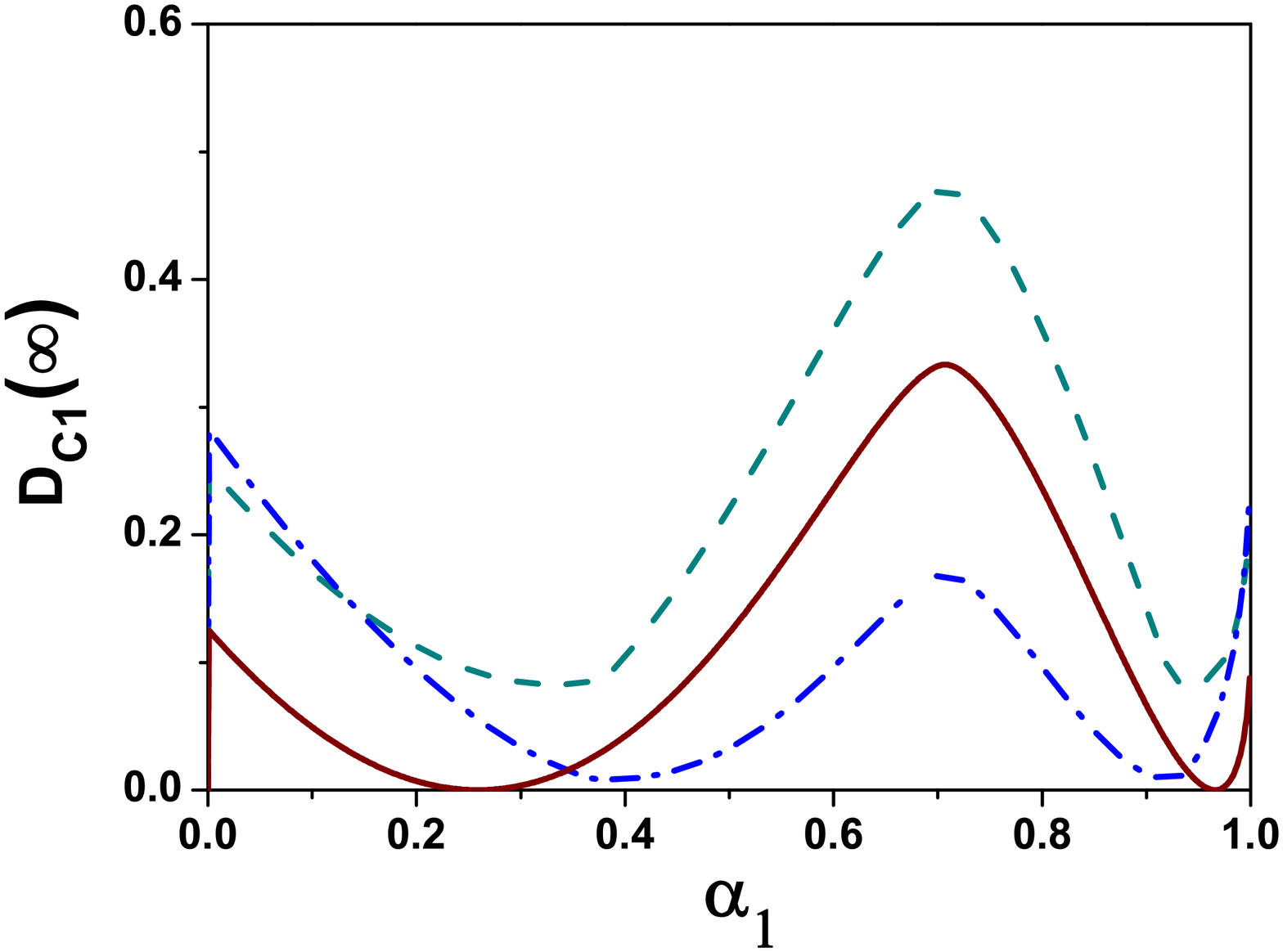}
 \caption{}
\end{subfigure}
\caption{ \small (a) Concurrence
 for the model $C2$ with $ \alpha_2=0.67$ and various parameters ($E$ and $\gamma$) of the Lorenzian density 
 of states of the environment (\ref{ldos}).
The green solid line with quite close $\tau_{ESD}$ and $\tau_{ESB}$ corresponds to $\gamma =0.15, E=1.1$,  the green arrow
indicates
 $ C_{C2}(\infty)=0.071$, i.e., the entanglement trapping. The red dash dot line corresponds to
  $\gamma =0.33,E=1.3$ with $0<\tau^{(1)}_{ESD}< \tau_{ESB}<\tau^{(2)}_{ESD} < \infty$.
The blue dash line corresponds to $ \gamma =0.33,\;E=1.5$  and displays the monotone decrease from the initial value
  to zero at $ \tau_{ESB}< \infty.$
 (b) Discord at infinity for the model $C1$ as a function of the entanglement parameter $\alpha_1$ in (\ref{1}) for various parameters $E$ and $\gamma$ of (\ref{ldos}). The green dash line corresponds to $\;\gamma = 0.2,\; E=0.5$, the blue dash dot line corresponds to $\;\gamma = 0.3,\; E=1.5$ and the brown  solid line corresponds to the
  "flat" density of states (\ref{fdos}). $D_{C1}( \infty)=0$ only in the last case and only
for $\alpha_1=0.5\sqrt{2\pm \sqrt{3}}$.
}
\end{figure}

Fig. 1(b). Unlike the models $I1$ and $F1$ of Fig. 1(a) displaying a single
ESD and no ESB, here, i.e., for the model $C1$ (common reservoir and the pure initial conditions (\ref{1}))  we have
multiple ESD's and ESB's. This is a manifestation of the
backaction of the environment in the non Markovian dynamics of entanglement,
resulting in our case from the indirect interaction (dynamical correlations) between the qubits via  
the common reservoir. Note the interplay between the behavior of $C_{C1}$ and $N_{C1}$: $C_{C1}>N_{C1}$ 
in the "life" periods and  $C_{C1}=N_{C1}=0$ in the "death" periods (cf. (\ref{c2n}) and (\ref{c2n0})) 
with the coinciding death and birth moments.
Passing from the initial conditions
(\ref{1}) to the looking quite similar initial condition (\ref{2}), we get a different behavior of the 
concurrence $C_{C2}$. Here we have just one ESD and one ESB with the subsequent positive 
values up to a certain positive value of $C_{C1}$ at infinity.
This behavior is known as the entanglement trapping, see e.g. \cite{Lo-Co:13}
for an analogous behavior in the model with a bosonic environment.

Fig. 2 shows the behavior of the concurrence for the cases where the initial state of two qubits can be unentangled (the initial state (\ref{0}) of Fig. 2(a) is unentangled for all $\alpha_0 \in [0,1]$ and the initial state (\ref{3}) of Fig. 2(b) is unentangled for $|\alpha_3|<1/3$). We see that in all these cases the entanglement is
absent during a certain initial period ($\min \tau_{ESD}=0$), then it appears at some $\tau_{ESB}>0$ and
displays the various types of behavior: fast and slow initial growth, multiple ESB's and ESD's  and subsequent decay and vanishing either at finite moment or at infinity (hence, the trapping again). The figure demonstrates the role of dynamical correlations between the qubits via the common environment in the "producing" of the entanglement. Note that for the models of independent qubits with Hamiltonians
(\ref{hi1}) and (\ref{hf1}), hence, without dynamical correlations, and with the same initial conditions ((\ref{0}) or (\ref{3})) the concurrence is identically zero, i.e., the entanglement is absent \cite{Br-Pa:18}. The same is true for certain bosonic environment \cite{Lo-Co:13,Ma-Co:09}.

Fig. 3 demonstrates the role of the density of states of reservoir (the Lorentzian (\ref{ldos}) in our case). It follows from Fig. 3(a) that by varying the parameters ($E,\gamma$) of the density of states, we can obtain the behavior similar that on Fig. 1(a) (red solid line), Fig. 1(b) (the blue solid line) and a "new" behavior (the green solid line) with the very close $\tau_{ESB}$ and $\tau_{ESB}$ resembling a cusp in the time scale of the figure.
On the other hand, according to Fig. 3(b), the behavior of the quantum discord at infinity as a function of the entanglement parameter $\alpha_1$ in (\ref{1}) is qualitatively similar for all considered values of ($E,\gamma$). There are, however, two special points $\alpha_1=\sqrt{2 \pm \sqrt{3}}/2$ where the discord is zero. It is widely believed that the cases where the discord vanishes are rather rare comparing with those for the concurrence \cite{Be-Co:17,Lo-Co:13}. In our case this happens for the indicated values of $E,\gamma$ and $\alpha_1$
and for the flat density of states (\ref{fdos}), where the corresponding reduced  density matrix is the "uniform" state
  $\rho(E,\infty )= 4^{-1} \mathrm{diag} (1,1,1,1)$
for which the discord is zero.
%

\begin{figure}[t!]
\begin{subfigure}[b]{.49\textwidth}
  \centering
  \includegraphics[width=\linewidth]{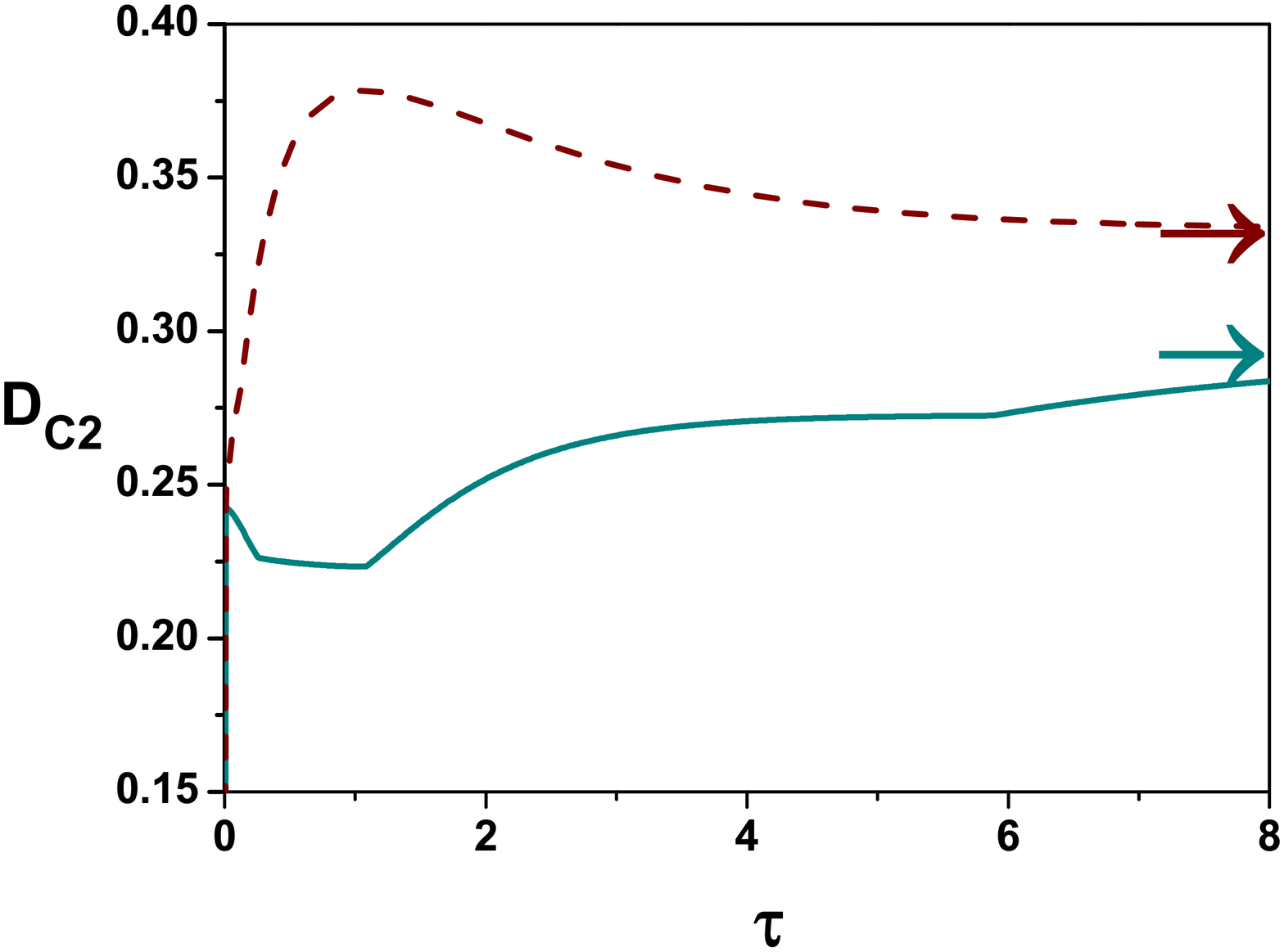}
 \caption{}
\end{subfigure}
\begin{subfigure}[b]{.49\textwidth}
  \centering
  \includegraphics[width=\linewidth] {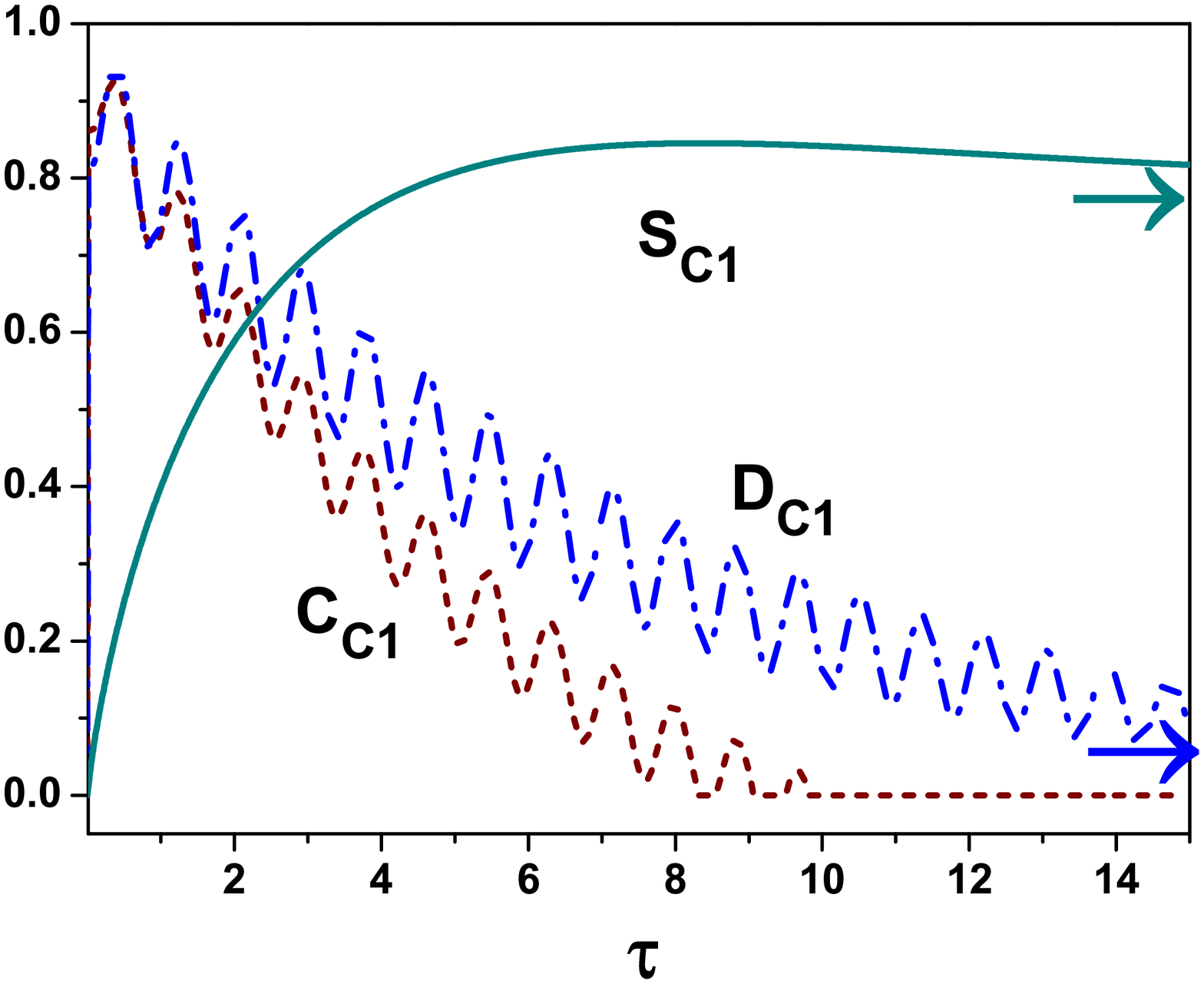}
 \caption{}
\end{subfigure}
\caption{ \small (a) The discord $D_{C2}$ as a function of $\tau$
for $\alpha_2=0.2$ where $D_{C2}(0)=0.242$. The solid green line
is $D_{C2}$ with $\gamma =0.8, E=2$, the green arrow indicates $
D_{C2}(\infty)=0.295$. The brown dash line is $D_{C2}$ for the
flat density of states (\ref{fdos}), the brown
 arrow  indicates $ D_{C2}( \infty)=0.333$.
 Both curves are not monotone.
(b)  The model $C1$ with $\alpha_1=1/2$.
The brown short dash line is the concurrence $C_{C1}, \; C_{C1}(0)=0.866,\; C_{C1}(\infty)=0$, the blue  dash dot line is the discord $D_{C1},\;D_{C1}(0)=0.811,\; D_{C1}(\infty)=0.039$ (blue arrow, the discord freezing) and
the green solid line is the entropy $S_{C1},\;S_{C1}(0)=0,\; S_{C1}(\infty)=0.791$ (green arrow). The concurrence decreases with
oscillations to zero at a certain moment (ESD) with two ESB and
two ESD in between. 
 }
\end{figure}

Fig. 4(a) illustrates the mentioned above point on the rarity of cases where the discord vanishes. Here the discord never 
vanishes and the lower plot is rather structured, containing, in particular, an almost flat segment, known as the discord 
freezing \cite{Be-Co:17,Lo-Co:13}. Fig. 4(b) displays the three discussed in Section 2.4 quantifiers of quantum 
coherence (the concurrence, the quantum discord and the entropy) with a quite structured behavior. Here the discord 
decreases with oscillations tending to a non zero value at infinity.
The concurrence oscillate as well with similar amplitude and frequency but becomes zero at a finite moment 
with several ESD and ESB before. Note that the corresponding value of $\alpha_1$ is 1/2, but it can be shown that
the analogous behavior of the concurrence holds for all $\alpha_1 \in (0,1)$ except $\alpha_1=2^{-1/2}$, where 
the concurrence is monotone. The entropy varies regularly from zero at zero (pure initial state) to a non zero value at infinity
but is not monotone. This has to be compared with the exact results of \cite{Lo-Co:13,Ma-Co:10} obtained for a 
particular solution of the spin boson model for two qubits with the common environment, where the entropy of the 
model also oscillates in time and the amplitude of oscillations is even considerably larger than that of the concurrence.

\section{ Large-$N$ behavior of the general reduced density matrix}

In this section we will prove a general version of our basic formulas (\ref%
{sa}) and (\ref{rhosg1}) -- (\ref{gz}). Namely, we consider a $pN\times pN$
analog
\begin{equation}
H_{\mathcal{S\cup E}}=H_{\mathcal{S}}\otimes \mathbf{1}_{\mathcal{E}}+%
\mathbf{1}_{\mathcal{S}}\otimes M_{N}+Q_{\mathcal{S}}\otimes W_{N},
\label{hp}
\end{equation}%
of the $4N\times 4N$ Hamiltonian (\ref{hc1}), where now $H_{\mathcal{S}}$
and $Q_{\mathcal{S}}$ are $p\times p$ hermitian matrices not necessarily
given by (\ref{hqub}) and (\ref{qs}) for $p=4$. We note that the
corresponding assertions as well as their proofs are generalizations of
those for the deformed semicircle law (DSCL) of random matrix theory, see
\cite{Pa-Sh:11}, Sections 2.2 and 18.3.

We will use the Greek indices varying from $1$ to $p$ to label the states of
the systems and the Latin indices varying from $1$ to $N$ to label the
states of the environment. Besides, we will not indicate as a rule the
dependence on $N$ of the many matrices below. Hence, we write the$\ pN\times pN$
and $p\times p$ density matrices of (\ref{dmg}) and (\ref{rdm}) as%
\begin{equation}
\rho _{\mathcal{S\cup E}}(t)=\{\rho _{\alpha j,\beta k}(t)\}_{\alpha ,\beta
=1,j,k=1}^{p,N},  \label{rhoge}
\end{equation}%
\begin{equation}
\;\rho _{\mathcal{S}}(t)=\{\rho _{\alpha \beta }(t)\}_{\alpha ,\beta
=1}^{p},\;\; \rho _{\alpha \beta }(t)=N^{-1}\sum_{j=1}^{N}\rho _{\alpha j,\beta
j}(t),\;\alpha ,\beta =1,...,p.  \label{rdmg}
\end{equation}%
Since the probability law (\ref{lGOE}) is unitary invariant, we can assume
without loss of generality that the hermitian matrix $M_{N}$ (the
environment Hamiltonian) in (\ref{hp}) is diagonal%
\begin{equation}
M_{N}=\{\delta _{jk}E_{N}^{(k)}\}_{j,k=1}^{N},  \label{mndi}
\end{equation}%
i.e., we can use the orthonormal basis of its eigenvectors as the basis in
the state space of the environment.

It follows then from (\ref{dmg}) -- (\ref{rdm}) and (\ref{psik}) that the
matrix form of the channel superoperator (\ref{super}) is
\begin{equation}
(\rho _{\mathcal{S}}(t))_{\alpha \beta }=\sum_{\gamma ,\delta =1}^{p}\Phi
_{\alpha \beta \gamma \delta }^{(k)}(t)(\rho _{\mathcal{S}}(0))_{\gamma
\delta },  \label{rFr0}
\end{equation}%
where%
\begin{equation}  \label{mef}
\Phi _{\alpha \beta \gamma \delta }^{(k)}(t)=\sum_{j=1}^{N}U_{\alpha
j,\gamma k}(-t)U_{\delta k,\beta j}(t),
\end{equation}%
and
\begin{equation}
U(\pm t)=e^{\pm itH}=\{U_{\alpha j,\beta k}(\pm t)\}_{\alpha ,\beta
=1,j,k=1}^{p,N}.  \label{uexp}
\end{equation}

\medskip \noindent 
\textbf{Result 1.} \emph{Consider the Hamiltonian (\ref{hp}) where $H_{%
\mathcal{S}}$ and $Q_{\mathcal{S}}$ are $p\times p$ arbitrary $N$%
-independent hermitian matrices, $M_{N}$ is a hermitian $N \times N$ matrix satisfying (\ref{non}) and (\ref{ele}%
) and $W_{N}$ is given by (\ref{lGOE}) -- (\ref{GOE}). Then we have for the
entries (\ref{rdmg}) of the reduced density matrix (\ref{rdmg})%
\begin{align}
&\mathbf{Var}\{\rho _{\alpha \beta }(t)\}  = \mathbf{E}\{|\rho _{\alpha
\beta }(t)|^{2}\}-|\mathbf{E}\{\rho _{\alpha \beta }(t)\}|^{2}  \notag \\
&\hspace{+2.5cm}\leq Ct^{2}\mathrm{Tr} Q_{\mathcal{S}}^{2}/N,\;C=4p^{2},\;\alpha ,\beta =1,...,p.
\label{sag}
\end{align}%
} 
\textbf{Proof of Result 1.} We view every $\rho _{\alpha \beta }(t)$ of (\ref{rFr0}) -- (\ref{uexp}) as a function of
the Gaussian random variables $\{W_{ab}\}_{a,b=1}^{N}$ of (\ref{lGOE}) -- (%
\ref{GOE}) and use the Poincar\'{e} inequality (see \cite{Pa-Sh:11},
Proposition 2.1.6) according to which we have for any differentiable and
polynomially bounded function $\varphi $ of the collection $%
\{W_{ab}\}_{a,b=1}^{N}$:%
\begin{equation}
\mathbf{Var}\{\varphi \}=\mathbf{E}\left\{ \left\vert \varphi \right\vert
^{2}\right\} -|\mathbf{E}\left\{ \varphi \right\} |^{2}\leq
N^{-1}\sum_{a,b=1}^{N}\mathbf{E}\left\{ \left\vert \frac{\partial \varphi }{%
\partial W_{ab}}\right\vert ^{2}\right\} .  \label{pob}
\end{equation}%
To find the derivatives of $\varphi =\rho _{\alpha \beta }(t)$ with respect
to $W_{ab}$, we use the Duhamel formula
\begin{equation}
\frac{d}{dx}e^{A(x)}=\int_{0}^{1}dse^{(1-s)A(x)}\frac{d}{dx}A(x)e^{sA(x)}
\label{difu}
\end{equation}%
valid for any differentiable matrix-function $A$ of $x$. In our case $A=itH_{\mathcal{A} \cup \mathcal{E}}$
of (\ref{hp}) viewed as a function of $W_{ab}$ for a given pair $(a,b)$. By
using (\ref{hp}), (\ref{uexp}), (\ref{pob}) and (\ref{difu}), we obtain
\begin{equation}
\frac{\partial }{\partial W_{ab}}U_{\alpha j,\beta k}(\pm t)=\pm
i\int_{0}^{t}ds\sum_{\alpha ^{\prime },\beta ^{\prime }=1}^{p}U_{\alpha
j,\alpha ^{\prime }a}(\pm (t-s))Q_{\alpha ^{\prime }\beta ^{\prime
}}U_{\beta ^{\prime }b,\beta k}(\pm s).  \label{dudw}
\end{equation}%
and we omit the subindex $\mathcal{S}$\thinspace\ in $Q$ here and often below.
This, (\ref{dmg}), (\ref{rdm}) and (\ref{hp}) yield%
\begin{equation}
\frac{\partial }{\partial W_{ab}}\rho _{\alpha \beta }(t)=T^{(1)}_{ab}+T^{(2)}_{ab},
\label{drtt}
\end{equation}%
where%
\begin{eqnarray}
T_{ab}^{(1)} &=&-\int_{0}^{t}ds\sum_{\alpha ^{\prime },\gamma ,\gamma
^{\prime },\delta =1}^{p}\sum_{j=1}^{N}U_{\alpha j,\alpha ^{\prime
}a}(-(t-s))Q_{\alpha ^{\prime }\gamma ^{\prime }}U_{\gamma ^{\prime
}b,\gamma k}(-s)\rho _{\gamma \delta }(0)U_{\delta k,\beta j}(t),  \notag \\
T_{ab}^{(2)} &=&-\int_{0}^{t}ds\sum_{\beta ^{\prime },\gamma ,\delta ,\delta
^{\prime }=1}^{p}\sum_{j=1}^{N}U_{\alpha j,\gamma k}(-t)\rho _{\gamma \delta
}(0)U_{\delta k,\delta ^{\prime }a}(t-s)Q_{\delta ^{\prime }\beta ^{\prime
}}U_{\beta ^{\prime }b,\beta j}(s).  \label{t1t2}
\end{eqnarray}%
We have then by (\ref{pob})%
\begin{eqnarray}
\mathbf{Var}\{\rho _{\alpha \beta }(t)\} &\leq &N^{-1}\sum_{a,b=1}^{N}%
\mathbf{E}\left\{ \left\vert T_{ab}^{(1)}+T_{ab}^{(2)}\right\vert
^{2}\right\}   \notag \\
&\leq &2N^{-1}\mathbf{E}\left\{ \sum_{a,b=1}^{N}\left\vert
T_{ab}^{(1)}\right\vert ^{2}\right\} +2N^{-1}\mathbf{E}\left\{
\sum_{a,b=1}^{N}\left\vert T_{ab}^{(2)}\right\vert ^{2}\right\} .
\label{var1}
\end{eqnarray}%
The first sum on the right of (\ref{var1}) is by (\ref{t1t2})
\begin{eqnarray*}
\sum_{a,b=1}^{N}\left\vert T_{ab}^{(1)}\right\vert ^{2} &=&\sum_{a,b=1}^{N}
\\
&&\hspace{-1cm}\times \left\vert \int_{0}^{t}ds\sum_{\alpha ^{\prime },\gamma ,\gamma
^{\prime },\delta =1}^{p}\sum_{j=1}^{N}U_{\alpha j,\alpha ^{\prime
}a}(-t+s)Q_{\alpha ^{\prime }\gamma ^{\prime }}U_{\gamma ^{\prime }b,\gamma
k}(-s)\rho _{\gamma \delta }(0)U_{\delta k,\beta j}(t)\right\vert ^{2}.
\end{eqnarray*}%
By applying Schwarz inequality to the sum over $\alpha ^{\prime },\gamma
,\gamma ^{\prime },\delta $ and to the integral over $t$,
we obtain
\begin{eqnarray}
\sum_{a,b=1}^{N}\left\vert T_{ab}^{(1)}\right\vert ^{2} &\leq
&t \sum_{\alpha',\gamma'=1}^p |Q_{\alpha' \gamma'}|^{2}\sum_{\gamma ,\delta =1}^{p}|\rho _{\gamma \delta
}(0)|^{2}\int_{0}^{t}ds\sum_{\gamma ^{\prime },b=1}^{p,N}|U_{\gamma ^{\prime
}b,\gamma k}(-s)|^{2}  \notag\\
&&\hspace{-2cm}\times \sum_{\gamma,\delta=1}^p \sum_{j_{1},j_{2}=1}^{N}\sum_{\alpha ^{\prime },a=1}^{p,N}U_{\alpha
j_{1},\alpha ^{\prime }a}(-(t-s))U_{\alpha j_{2},\alpha ^{\prime }a}^{\ast
}(-(t-s))U_{\delta k,\beta j_{1}}(t)U_{\delta k,\beta j_{2}}^{\ast }(t),
\label{tab1}
\end{eqnarray}%
where the symbol "$\ast $" denotes the complex conjugate.

Recalling now that $U(t)$ is unitary group, hence, $U_{\alpha j,\beta
k}^{\ast }(t)=U_{\beta k,\alpha j}^{\ast }(-t)$ and for any $\alpha
_{1},\alpha _{2}$, $j_{1},j_{2},s_{1},s_{2}$
\begin{eqnarray}
\sum_{\alpha \text{ }^{\prime },a=1}^{p,N}U_{\alpha _{1}j_{1},\alpha
^{\prime }a}(s_{1})U_{\alpha _{2}j_{2},\alpha ^{\prime }a}^{\ast }(s_{2})
&=&U_{\alpha _{1}j_{1},\alpha _{2}j_{2}}(s_{1}-s_{2}),\;  \label{unit1} \\
U_{\alpha _{1}j_{1},\alpha _{2}j_{2}}(0) &=&\delta _{\alpha _{1}\alpha
_{2}}\delta _{j_{1}j_{2}},  \notag
\end{eqnarray}%
we obtain that the sum over $(\alpha',\gamma')$ is $\mathrm{Tr} Q^2$,
the first sum over $(\gamma ,\delta )$ is $\mathrm{Tr}\rho
^{2}(0)\leq 1$,  the sum over ($\gamma ^{\prime },b$) \
is 1 by (\ref{unit1}), the sum over ($\alpha ^{\prime },a)$ is $\delta _{j_{1}j_{2}}$ again
by (\ref{unit1}) and then the sum over $j=j_{1}=j_{2}$ is bounded by 1 also by (\ref{unit1}) and the second sum over $(\gamma,\delta)$ is $p^2$. We conclude that
\begin{equation*}
\sum_{a,b=1}^{N}\left\vert T_{ab}^{(1)}\right\vert ^{2}\leq
p^{2}t^{2}\mathrm{Tr}_{\mathcal{S}} Q^2_{\mathcal{S}}.
\end{equation*}%
An analogous argument yields the same bound for the second sum in (\ref{var1}%
) and we obtain (\ref{sag}). $\blacksquare $ 

\medskip\noindent
\textbf{Result 2.} \emph{In the setting of Result 1 above we have uniformly in $t$ varying on any
compact interval of $[0,\infty )$:
\begin{eqnarray}
\rho (E,t) &=&\lim_{N\rightarrow \infty }\mathbf{E}\{\rho _{S}(t)\}  \notag
\\
&=&-\frac{1}{(2\pi i)^{2}}\int_{-\infty -i\varepsilon }^{\infty
-i\varepsilon }dz_{1}\int_{-\infty +i\varepsilon }^{\infty +i\varepsilon
}dz_{2}e^{i(z_{1}-z_{2})t}F(E,z_{1},z_{2}),  \label{rhosg1}
\end{eqnarray}%
where $F$ solves the linear $p\times p$ matrix equation
\begin{align}
& F(E,z_{1},z_{2})=F_{0}(E,z_{1},z_{2})\notag
\\& +\int_{-\infty }^{\infty }G(E^{\prime },z_{2})Q_{\mathcal{S}%
}F(E,z_{1},z_{2})Q_{\mathcal{S}}G(E^{\prime },z_{1})\nu _{0}(E^{\prime
})dE^{\prime }  \label{brevG1}
\end{align}%
with the density of states of the environment $\nu_0$ defined in (\ref{non}),
\begin{equation}
F_{0}(E,z_{1},z_{2})=G(E,z_{2})\rho _{\mathcal{S}}(0)G(E,z_{1}),  \label{fo1}
\end{equation}%
\begin{equation}
G(E,z)=(E+H_{\mathcal{S}}-z-Q_{\mathcal{S}}\mathsf{G}(z)Q_{\mathcal{S}})^{-1}
\label{gez1}
\end{equation}%
and $\mathsf{G}(z)$ solving uniquely the non-linear $p\times p$ matrix equation
\begin{equation}
\mathsf{G}(z)=\int_{-\infty }^{\infty }(E^{\prime }+H_{\mathcal{S}}-z-Q_{%
\mathcal{S}}\mathsf{G}(z)Q_{\mathcal{S}})^{-1}\nu _{0}(E^{\prime
})dE^{\prime } \label{gz1}
\end{equation}
in the class of $p\times p$ analytic in $%
\mathbb{C}\setminus \mathbb{R}$ matrix functions such that
\begin{equation}
\Im \mathsf{G}(z)\Im z>0,\;\Im z\neq 0,\;\sup_{y\geq 1}y||\mathsf{G}%
(iy)||\leq \infty. \label{neva}
\end{equation}%
}

\noindent
\textbf{Proof of Result 2.}
This proof is more involved than that of Result 1. We will start with the
asymptotic analysis of
\begin{equation}
\overline{U}_{\alpha j,\beta k}(t)=\mathbf{E}\{U_{\alpha j,\beta k}(t)\},
\label{ubar}
\end{equation}%
i.e., the first moment of the evolution operator (\ref{uexp}), since the moment is
necessary for the asymptotic analysis of the second moment, i.e., according
to (\ref{mef}), of%
\begin{equation}
\mathbf{E}\{\Phi _{\alpha \beta \gamma \delta }^{(k)}(t)\}=\mathbf{E}\left\{
\sum_{j=1}^{N}U_{\alpha j,\gamma k}(-t)\}U_{\delta k,\beta j}(t)\}\right\} .
\label{eF}
\end{equation}
which results in (\ref{brevG1}) -- (\ref{gz1}). Besides, the asymptotic
analysis of (\ref{ubar}) includes several important technical steps which
are also used in the analysis of (\ref{eF}), but are less tedious and more
transparent for (\ref{ubar}) than for (\ref{eF}).

\textit{Asymptotic analysis of} (\ref{ubar}). It is convenient to pass
from the evolution operator (\ref{uexp}) of the $pN\times pN$ hermitian
matrix $H_{\mathcal{S\cup E}}$ of (\ref{hp}) to its resolvent
\begin{equation}
G_{H_{\mathcal{S\cup E}}}(z)=(H_{\mathcal{S\cup E}}-z)^{-1}=\{G_{\alpha
j,\beta k}(z)\}_{\alpha ,\beta ,j,k=1}^{pN},\;\Im z\neq 0  \label{gse}
\end{equation}%
by using the formulas%
\begin{equation}
G_{H_{\mathcal{S\cup E}}}(z)=\pm i\int_{0}^{\infty }dte^{\mp itz}U_{H_{%
\mathcal{S\cup E}}}(\pm t),\;\Im z\lessgtr 0.  \label{lap}
\end{equation}%
Given the resolvent we obtain the evolution operator via the inversion
formula%
\begin{equation}
U_{H_{\mathcal{S\cup E}}}(\pm t)=\mp \frac{1}{2\pi i}\int_{-\infty
\pm i\varepsilon}^{\infty
\pm i\varepsilon}dt e^{\pm itz}G_{H_{\mathcal{S\cup E}}}(z),  \label{ilap}
\end{equation}%
where the integral is understood in the Cauchy sense at infinity.

In view of the above formulas is suffices to find an asymptotic form of the
expectation (the first moment)
\begin{equation}
\overline{G}_{H_{\mathcal{S\cup E}}}(z)=\mathbf{E}\{G_{H_{\mathcal{S\cup E}%
}}(z)\}=\{\overline{G}_{\alpha j,\beta k}(z)\}_{\alpha ,\beta
=1,j,k=1}^{p,N},\;\Im z\neq 0.  \label{cgbar}
\end{equation}%
of the resolvent (\ref{gse}).

To this end we will use an extension of the tools of random matrix theory as
they presented in \cite{Pa-Sh:11} and used there to derive the so called
deformed semicircle law for Gaussian random matrices, the "scalar" case of $p=1$ and then, in \cite%
{Le-Pa:03}, to deal with the one-qubit case of $p=2$ for (%
\ref{hp}).

Denote for brevity%
\begin{eqnarray}
&& H_{\mathcal{S\cup E}} =H=H^{(0)}+H^{(1)},\;  \notag \\
&& H^{(0)} =H_{\mathcal{S}}\times 1_{\mathcal{E}}+1_{\mathcal{S}}\times
M_{N},\;H_{1}=H_{\mathcal{SE}}=vQ_{\mathcal{S}}\times W_{N}  \label{hs}
\end{eqnarray}%
in (\ref{hp}) (recall that $H_{\mathcal{S}}$ and $Q_{\mathcal{S}}$ are now
arbitrary $p\times p$ hermitian matrices, i.e., not necessarily given by (%
\ref{hqub}) and (\ref{qs})). Set%
\begin{equation*}
G(z)=(H-z)^{-1},\;G^{(0)}(z)=(H^{(0)}-z)^{-1},\;\Im z\neq 0
\end{equation*}%
and use the resolvent identity
\begin{equation}
G(z)=G^{(0)}(z)-G(z)H^{(1)}G^{(0)}(z)  \label{resid}
\end{equation}%
to write%
\begin{equation}
\overline{G}_{\alpha j,\beta k}=G_{\alpha j,\beta k}^{(0)}-\sum_{\alpha
^{\prime },\beta ^{\prime }=1,}^{p}\sum_{j',k'=1}^{N}\mathbf{E}%
\{G_{\alpha j,\alpha
^{\prime }j^{\prime }}Q_{\alpha ^{\prime }\beta ^{\prime }}W_{j^{\prime }k^{\prime }}\}G_{\beta ^{\prime }k^{\prime },\beta k}^{(0)},
\label{cgb0}
\end{equation}%
where we omit the subindex $\mathcal{S}$ in $Q_{\mathcal{S}}$ and the
argument $z$ in $G$.

To proceed we will use the Gaussian differentiation formula, according to
which if $\{W_{ab}\}_{a,b=1}^{N}$ is the collection of complex Gaussian
random variables (\ref{lGOE}) -- (\ref{GOE}) and $\varphi $ is a
differentiable and polynomially bounded function of the collection, then
(see \cite{Pa-Sh:11}, Section 2.1)%
\begin{equation}
\mathbf{E}\{W_{ab}\varphi \}=N^{-1}\mathbf{E}\left\{ \frac{\partial \varphi
}{\partial W_{ba}}\right\} ,\;a,b=1,...,N.  \label{difgc}
\end{equation}%
Viewing $G_{\alpha j,\alpha^{\prime } j^{\prime }}$ as function of a particular $%
W_{ab}$ and using the formula (cf. (\ref{dudw})),
\begin{equation}
\frac{\partial }{\partial W_{ab}}G_{\rho r,\sigma s}=-\sum_{\rho ^{\prime
},\sigma ^{\prime }=1}^{p}G_{\rho r,\rho ^{\prime }a}Q_{\rho ^{\prime
}\sigma ^{\prime }}G_{\sigma ^{\prime }b,\sigma s},  \label{dcgdw}
\end{equation}%
which follows easily from the resolvent identity (\ref{resid}) (cf. (\ref%
{difu})), we obtain from (\ref{cgb0})%
\begin{eqnarray}
\overline{G}_{\alpha j ,\beta k} &=&G_{\alpha j,\beta k}^{(0)}-N^{-1}%
\sum_{\alpha ^{\prime },\beta ^{\prime }=1,j^{\prime },k^{\prime }=1}^{p,N}%
\mathbf{E}\left\{ \frac{\partial }{\partial W_{k^{\prime }j^{\prime }}}%
G_{\alpha j,\alpha ^{\prime }j^{\prime }}\right\} Q_{\alpha ^{\prime }\beta
^{\prime }}G_{\beta ^{\prime }k^{\prime },\beta k}^{(0)}  \notag \\
&=&G_{\alpha j,\beta k}^{(0)}+\sum_{\alpha ^{\prime },\beta ^{\prime }=1}^{p}%
\sum_{j'=1}^N \mathbf{E}\left\{ G_{\alpha j,\alpha ^{\prime }j^{\prime }}(g_{Q})_{\alpha
^{\prime }\beta ^{\prime }}\right\} G_{\beta ^{\prime }j^{\prime },\beta
k}^{(0)},  \label{cgb1}
\end{eqnarray}%
where%
\begin{eqnarray}
g_{Q}=QgQ, \; \; g &=&N^{-1}\mathrm{Tr}_{\mathcal{E}}G=\{g_{\gamma \delta
}\}_{\gamma ,\delta =1}^{p},\;  \notag \\
g_{\gamma \delta} &=&N^{-1}\sum_{k^{\prime }=1}^{N}G_{\gamma k^{\prime
},\delta k^{\prime }}.  \label{gqg}
\end{eqnarray}%
Writing
\begin{equation}
g=\overline{g}+g^{\circ },\;\mathbf{E}\left\{ g^{\circ }\right\} =0,
\label{4g}
\end{equation}%
we present (\ref{cgb1}) in the compact matrix form%
\begin{equation}
\overline{G}=G^{(0)}+\overline{G}(\overline{g}_{Q}\times \mathbf{1}_{%
\mathcal{E}})G^{(0)}+R^{(1)}G^{(0)}  \label{cgb2}
\end{equation}%
with%
\begin{equation}
R^{(1)}=\mathbf{E}\left\{ G  (g_{Q}^{\circ }\times \mathbf{1}_{\mathcal{E}%
})\right\} .  \label{cre0}
\end{equation}%
Denote $\{\lambda _{\tau t}\}_{\tau,t=1}^{pN}$ and $\{\Psi _{\tau
t}\}_{\tau,t=1}^{pN}$ the eigenvalues (possibly repeating) and
orthonormal eigenvectors of $pN\times pN$ matrix $H$ of (\ref{hs}). It
follows then from the spectral theorem that%
\begin{equation}
G=(H-z)^{-1}=\sum_{\tau ,t=1}^{pN}\frac{1}{\lambda _{\tau t}-z}P_{\tau t},
\label{gspec}
\end{equation}%
where $P_{\tau t}$ is the orthogonal projection on $\Psi _{\tau t}$.

Writing further for any matrix $A$%
\begin{equation}\label{ima}
\Im A:=(A-A^{+})/2i,
\end{equation}%
where $A^{+}$ is the hermitian conjugate of $A$ and for any hermitian
matrices $A$ and $B$%
\begin{equation*}
A>B
\end{equation*}%
if $A-B$ is positive definite, we have from (\ref{gspec})%
\begin{equation*}
\Im G(z)/\Im z>0,\;\Im z\neq 0.
\end{equation*}%
The same inequality holds for $\overline{G}$, $\overline{g}=N^{-1}\mathrm{Tr}%
_{\mathcal{E}}\overline{G}$ and $\overline{g}_{Q}$ of (\ref{gqg}). This
implies the bound%
\begin{equation}\label{imhg}
\Im (H^{(0)}-z-\overline{g}_{Q})/\Im z=-(\mathbf{1}_{\mathcal{S}}+\Im Q%
\overline{g}Q/\Im z)\times \mathbf{1}_{\mathcal{E}}<-\eta \mathbf{1}_{%
\mathcal{S\cup E}},
\end{equation}%
with an $N$-independent $\eta >0$ and%
\begin{equation}
|\Im z|\geq \eta >0.  \label{ize}
\end{equation}%
Hence, the $pN\times pN$ matrix $H^{(0)}-z-\overline{g}_{Q}(z)\times \mathbf{%
1}_{\mathcal{E}}$ is invertible uniformly in $N$ and (\ref{cgb2}) is
equivalent to%
\begin{equation}
\overline{G}(z)=G^{(0)}(z_Q)+R^{(1)}(z)G^{(0)}(z_{Q}),\;z_{Q}=z%
\mathbf{1}_{\mathcal{S}}+\overline{g}_{Q}(z).  \label{cgb3}
\end{equation}%
Note now that $H^{(0)}$ of (\ref{hs}) admits the separation of variables,
hence, in its spectral representation (see (\ref{gspec})) $\lambda _{\tau
t}=\varepsilon _{\tau }+E_{t}$, $\Psi _{\tau t}=\psi _{\tau }\otimes \Psi
_{t}$, where $\{\varepsilon _{\tau }\}_{\tau =1}^{p}$ and $%
\{E_{t}\}_{t=1}^{N}$ are the eigenvalues $\ $and $\{\psi _{\tau }\}_{\tau
=1}^{p}$ and $\{\Psi _{t}\}_{t=1}^{N}$\ are the eigenvectors of $H_{\mathcal{%
S}}$ and $H_{\mathcal{E}}=M_{N}$. Besides, $M_{N}$ is diagonal, see (\ref%
{mndi}), hence, $\Psi _{t}=\{\delta _{jt}\}_{j=1}^{N}$ and we have from (\ref%
{gspec})%
\begin{eqnarray}
G_{\alpha j,\beta k}^{(0)}(z) &=&\delta _{jk}G_{\alpha \beta }^{\mathcal{S}%
}(z-E_{k}),\;  \notag \\
G^{\mathcal{S}}(z) &=&(H_{\mathcal{S}}-z)^{-1}=\{G_{\alpha \beta }^{\mathcal{%
S}}(z)\}_{\alpha ,\beta =1}^{p}.  \label{cg0k}
\end{eqnarray}%
This allows us to write (\ref{cgb3}) as%
\begin{equation}
\overline{G}_{\alpha j,\beta k}(z)=\delta _{jk}G_{\alpha \beta }^{\mathcal{S}%
}(z_{Q}-E_{j})+\sum_{\gamma =1}^{p}R^{(1)}_{\alpha j,\gamma k}G_{\gamma \beta }^{%
\mathcal{S}}(z_{Q}-E_{j}),  \label{cgb4}
\end{equation}%
and, combining (\ref{non}), (\ref{gqg}) and (\ref{cgb4}), we get%
\begin{equation}
\overline{g}(z)=\int G^{\mathcal{S}}(z_{Q}-E)\nu _{N}(E)dE+r(z),
\label{eqg0}
\end{equation}%
where%
\begin{equation}
r_{\alpha \beta }(z)=N^{-1}\sum_{\gamma =1}^{p}\sum_{k=1}^{N}\mathbf{E}%
\left\{ G_{\mathcal{\alpha }k,\gamma k}(z)((Qg^{\circ
}Q)G^{\mathcal{S}}(z_{Q}-E_{k}))_{\gamma \beta }\right\} .  \label{rab}
\end{equation}%
It follows from (\ref{gspec}) and (\ref{imhg}) that%
\begin{equation}
|G_{j\alpha ,j\beta }(z)|\leq ||G(z)||\leq |\Im
z|^{-1},\;||G^{\mathcal{S}}(z_{Q}-E_{k})||\leq |\Im z|^{-1}.  \label{goc}
\end{equation}%
This and the bound
\begin{equation}\label{noqu}
|Q_{\alpha,\beta}| \le ||Q||, \; \alpha, \beta=1,...,p
\end{equation}
imply%
\begin{equation}
|r_{\alpha \beta }(z)|\leq \frac{p^{2}||Q||^{2}}{|\Im z|^{2}}\sum_{\gamma
,\delta =1}^{p}\mathbf{E\{}|g_{\gamma \delta }^{\circ }|\}\leq \frac{%
p^{2}||Q||^{2}}{|\Im z|^{2}}\sum_{\gamma ,\delta =1}^{p}\mathbf{Var}%
^{1/2}\left\{ g_{\gamma \delta }\right\} .  \label{rab1}
\end{equation}
We will bound $\mathbf{Var}\left\{ g_{\gamma \delta }\right\} $ by using
again the Poincar\'{e} inequality (\ref{pob}). We have by (\ref{dcgdw})
\begin{eqnarray}
\frac{\partial g_{\gamma \delta }}{\partial W_{ab}} &=&-\frac{1}{N}%
\sum_{j=1}^{N}\sum_{\gamma ^{\prime },\delta ^{\prime }=1}^{p}G_{\gamma
j,\gamma ^{\prime }a}G_{\delta ^{\prime }b,\delta j}Q_{\gamma ^{\prime
}\delta ^{\prime }}  \notag \\
&=&-\frac{1}{N}\sum_{\gamma ^{\prime },\delta ^{\prime }=1}^{p}Q_{\gamma
^{\prime }\delta ^{\prime }}\left( \sum_{j=1}^{N}G_{\gamma j,\gamma ^{\prime
}a}G_{\delta ^{\prime }b,\delta j}\right)   \label{dgabdw}
\end{eqnarray}
and then, by Schwarz inequality and (\ref{noqu})
\begin{eqnarray*}
\left\vert \frac{\partial g_{\gamma \delta }}{\partial W_{ab}}\right\vert
^{2} &\leq &\frac{p^2||Q||^2}{N^{2}}\sum_{\gamma ^{\prime },\delta
^{\prime }=1}^{p}\sum_{a,b=1}^{N}\left\vert \sum_{j=1}^{N}G_{\gamma j,\gamma
^{\prime }a}G_{\delta ^{\prime }b,\delta j}\right\vert ^{2}.
\end{eqnarray*}
Plugging this into the r.h.s. of (\ref{pob}) with $\varphi =g_{\gamma \delta
}$, we obtain
\begin{equation*}
\mathbf{Var}\left\{ g_{\gamma \delta }\right\} \leq \frac{p^2 ||Q||^{2}}{%
N^{3}}\mathrm{Tr}_{\mathcal{E}}\Gamma _{\gamma \gamma }\Gamma _{\delta
\delta }^{\ast },
\end{equation*}%
where
\begin{equation*}
\Gamma _{\alpha \alpha }=\{(GG^{+})_{\alpha j_{1},\alpha
j_{2}}\}_{j_{1},j_{2}=1}^{N}
\end{equation*}%
is the $N\times N$ matrix and it follows from (\ref{gspec}) that $||\Gamma
_{\alpha \alpha }||\leq ||\Im z|^{-2}$. This and the bound  $|\mathrm{Tr}_{%
\mathcal{E}}A|\leq ||A||N$ valid for any $N\times N$ matrix yield
\begin{equation}
\mathbf{Var}\{g_{\alpha \beta }(z)\}\leq \frac{p^{2}||Q||^{2}}{N^{2}|\Im
z|^{4}}  \label{vgoc}
\end{equation}%
implying together with (\ref{rab1})%
\begin{equation}
|r_{\alpha \beta }(z)|\leq \frac{p^{3}||Q||^{3}}{N|\Im z|^{4}},\;\alpha
,\beta =1,...,p.  \label{rab2}
\end{equation}%
The bound and the standard argument of random matrix theory (see \cite%
{Pa-Sh:11}, Chapter 2) allow us to conclude that the sequence $\{\overline{g}%
_{N}\}_{N}$ of $p\times p$ analytic in $\mathbb{C}\setminus \mathbb{R}$
matrix function (\ref{gqg}) contains a subsequence $\{\overline{g}%
_{N_{n}}\}_{n}$ which converges uniformly on any compact set of $\mathbb{C}%
\setminus \mathbb{R}$ to a unique solution $\mathsf{G}(z)$ of the matrix functional equation (\ref{gz1}) -- (\ref{neva}).
 Hence, the whole sequence $\{\overline{g}_{N}\}_{N}$ converges uniformly on
any compact set of $\mathbb{C}\setminus \mathbb{R}$ to the limit $\mathsf{G}$
solving uniquely (\ref{gz1}) -- (\ref{neva}).

Note that this assertion is a matrix analog of that on the so-called
deformed semicircle law of random matrix theory, see \cite{Pa-Sh:11},
Chapter 2. In articular, the proof of the unique solvability of (\ref{gz}) -- (\ref{neva}) repeats almost literally the corresponding proof in \cite{Pa-Sh:11}.

Consider now the expectation (\ref{cgbar}) of the resolvent. It it is easy
to see that a slightly modified version of an argument proving (\ref{rab2})
yields for the second term of (\ref{cgb4}) the bound coinciding with the
r.h.s. of (\ref{rab2}), i.e.,%
\begin{equation}
\left\vert \overline{G}_{\alpha j,\beta k}(z)-\delta _{jk}G_{\alpha \beta }^{%
\mathcal{S}}(z_{Q}(z)-E_{k})\right\vert \leq \frac{p^{3}||Q||^{3}}{N|\Im z|^{4}}%
,\;\alpha ,\beta =1,...,p,\;j=1,...,N.  \label{cgbf}
\end{equation}
This bound implies for any $z$ satisfying (\ref{ize})$\;$and all $\alpha
,\beta =1,...,p<\infty $%
\begin{equation}
\lim_{N\rightarrow \infty }\overline{G}_{\alpha j,\beta k}(z)=0,\;j\neq
k,\;j,k=1,...,N,   \label{lignd}
\end{equation}%
and if $k=k_{N}\rightarrow \infty ,\;N\rightarrow \infty $\ and is such that
(\ref{ele}) holds, then%
\begin{eqnarray}
G(E,z) &=&\lim_{N\rightarrow \infty }\overline{G}_{\alpha k_{N},\beta
k_{N}}(z)=G_{\alpha \beta }^{\mathcal{S}}(Z_{Q}-E), \notag \\
Z_{Q}(z) &=&z \mathbf{1}_{\mathcal{S}}+Q\mathsf{G}(z)Q,  \label{ligdi}
\end{eqnarray}%
where $Z_{Q}$ the $N\rightarrow \infty $  limit
of the $p \times p$ matrix function $z_{Q}$ given in (\ref{cgb3}).

Note now that by (\ref{uexp}) and (\ref{hs})
\begin{equation*}
\frac{d}{dt}\overline{U}_{\alpha j,\beta k}(t)=i\mathbf{E}\{(U(t)H)_{\alpha
j,\beta k}\}=i\sum_{\gamma ,l=1}^{p,N}\mathbf{E}\{U(t)_{\alpha j,\gamma
l}H_{\gamma l,\beta k}\},
\end{equation*}%
hence, by Schwarz inequality,%
\begin{equation*}
\left\vert \frac{d}{dt}\overline{U}_{\alpha j,\beta k}(t)\right\vert \leq
\mathbf{E}^{1/2}\left\{ \sum_{\gamma ,l=1}^{p,N}|U(t)_{\alpha j,\gamma
l}|^{2}\right\} \mathbf{E}^{1/2}\left\{ \sum_{\gamma ,l=1}^{p,N}|H_{\gamma
l,\beta k}|^{2}\right\} .
\end{equation*}%
The first factor on the right is bounded by 1 in view of (\ref{unit1}) and
according to (\ref{hs}) and (\ref{GOE}) the second factor admits the bound%
\begin{equation*}
3^{1/2}\left( \sum_{\gamma =1}^{p}|H_{\gamma \beta }^{\mathcal{S}%
}|^{2}+|E_{k}^{(N)}|^{2}+||Q||^{2}\sum_{l=1}^{N}\mathbf{E\{}%
|W_{lk}|\}^{2}\right) ^{1/2}
\end{equation*}%
It follows from (\ref{GOE}) that the above expression is bounded in $\alpha
,\beta ,j,k,N$ provided (\ref{ele}) is valid. Thus, the collection of continuous in $t$ functions $%
\overline{U}_{\alpha j,\beta k}:\mathbb{R}\rightarrow \mathbb{C}$ contains a
subsequence $\{\overline{U}_{\alpha j^{(N)},\beta k^{(N)}}\}_{N}$ in $N$
(where ($j_{N},k_{N}$) do not necessarily depend on $N$) which converges
uniformly in $t\in \lbrack 0,t_{0}],\;\forall t_{0}<\infty $ to a certain
continuous function. This, (\ref{lap}), (\ref{ilap}) and (\ref{lignd}) -- (%
\ref{ligdi}) imply for any $\alpha ,\beta =1,...,p $%
\begin{equation}
\lim_{N\rightarrow \infty }\overline{U}_{\alpha j,\beta k}(t)=0,\;j\neq
k,\;j,k=1,...,N,  \label{liund}
\end{equation}%
and if $k=k^{(N)}\rightarrow \infty ,\;N\rightarrow \infty $ is such that (%
\ref{ele}) holds, then%
\begin{equation}
\lim_{N\rightarrow \infty }\overline{U}_{\alpha k^{(N)},\beta k^{(N)}}(t)=%
\frac{1}{2\pi i}\int_{-\infty-i \varepsilon}^{\infty-\varepsilon}
e^{itz}G_{\alpha \beta }^{\mathcal{S}%
}(Z_{Q}(z)-E)dz,  \label{liud}
\end{equation}%
where $Z_{Q}$ is defined in (\ref{ligdi}).

\medskip
\textit{Asymptotic analysis of the channel operator.}
It follows from Result 1 above that it suffices to consider the expectation%
\begin{equation}
\overline{\Phi }_{\alpha \beta \gamma \delta }^{(k)}(t)=\sum_{j=1}^{N}%
\mathbf{E}\{U_{\alpha j,\gamma k}(-t)U_{\delta k,\beta j}(t)\}.
\label{chope}
\end{equation}%
of the entries (\ref{mef}) of the superoperator.

Introduce%
\begin{eqnarray}
\Phi _{\alpha \beta \gamma \delta }^{(k)}(t_{1},t_{2})
&=&\sum_{j=1}^{N}U_{\alpha j,\gamma k}(-t_{2})U_{\delta k,\beta
j}(t_{1}),\;t_{1}\geq 0,\;t_{2} \geq 0,  \notag \\
\Phi _{\alpha \beta \gamma \delta }^{(k)}(t) &=&\Phi _{\alpha \beta \gamma
\delta }^{(k)}(-t,t),\;t \geq 0,  \label{f2t}
\end{eqnarray}%
and pass from the evolution operator (\ref{uexp}) of the total hamiltonian $%
H_{\mathcal{S\cup E}}$ to its resolvents (\ref{gse}) by applying (\ref%
{lap}) with respect to $t_{1}$ and $t_{2}$. The result is%
\begin{equation}
F_{\alpha \beta \gamma \delta }^{(jk)}(z_{1},z_{2})=\sum_{j=1}^{N}G_{\alpha
j,\gamma k}(z_{2})G_{\delta k,\beta j}(z_{1}),\;\Im z_{1}<0,\;\Im z_{2}>0.
\label{cgaka}
\end{equation}%
with%
\begin{eqnarray}
\overline{F}_{\alpha \beta \gamma \delta }^{(jk)}(z_{1},z_{2}) &=&\mathbf{E}%
\{F_{\alpha \beta \gamma \delta }^{(jk)}(z_{1},z_{2})\},  \notag \\
\overline{F}_{\alpha \beta \gamma \delta }^{(k)}(z_{1},z_{2})
&=&\sum_{j=1}^{N}\mathbf{E}\{F_{\alpha \beta \gamma \delta
}^{(jk)}(z_{1},z_{2})\}.  \label{gakpk}
\end{eqnarray}%
We apply now to $\overline{F}_{\alpha \beta \gamma \delta
}^{(jk)}(z_{1},z_{2})$ the scheme of analysis analogous to that for (\ref%
{cgbar}). We use first the resolvent identity (\ref{resid}) for the second
factor $G_{\delta k,\beta j}(z_{1})$ on the right of (\ref{cgaka}) and then
the differentiation formulas (\ref{difgc}) and (\ref{dcgdw}). This yields (cf. (%
\ref{cgb1}))%
\begin{eqnarray}
\overline{F}_{\alpha \beta \gamma \delta }^{(jk)} &=&\overline{G}_{\alpha
j,\gamma k}(z_{2})G_{\delta k,\beta j}^{(0)}(z_{1})  \notag \\
&&\hspace{-1.5cm}+\sum_{\alpha ^{\prime },\beta ^{\prime }=1}^{p}\sum_{j^{\prime }=1}^{N}%
\mathbf{E}\{G_{\alpha j,\gamma k}(z_{2})G_{\delta k,\alpha ^{\prime
}j^{\prime }}(z_{1})(g_{Q}(z_{1}))_{\alpha ^{\prime }\beta ^{\prime
}}\}G_{\beta ^{\prime }j^{\prime },\beta j}^{(0)}(z_{1})  \notag \\
&&\hspace{-1.5cm}+N^{-1}\sum_{\alpha ^{\prime },\beta ^{\prime },\gamma ^{\prime}=1}^{p}\sum_{k^{\prime }=1}^{N}\mathbf{E}\{G_{\alpha j,\alpha ^{\prime
}j}(z_{2})Q_{\alpha ^{\prime }\gamma ^{\prime }}G_{\gamma ^{\prime
}k^{\prime },\gamma k}(z_{2})G_{\delta k,\delta ^{\prime }k^{\prime
}}(z_{1})\}Q_{\delta ^{\prime }\beta ^{\prime }}G_{\beta ^{\prime }j^{\prime
},\beta \gamma }^{(0)}(z_{1})  \label{cgab1}
\end{eqnarray}%
with $g_{Q}$ given by (\ref{gqg}) and then (\ref{cg0k}) implies%
\begin{eqnarray}
\overline{F}_{\alpha \beta \gamma \delta }^{(jk)} &=&\delta _{jk}\overline{G}%
_{\alpha j,\gamma k}(z_{2})G_{\delta \beta }^{\mathcal{S}}(z_{1}-E_{j})
\notag \\
&&\hspace{-1.5cm}+\sum_{\alpha ^{\prime },\beta ^{\prime }=1}^{p}\mathbf{E}\{F_{\alpha
\alpha ^{\prime }\gamma \alpha ^{\prime }}^{(jk)}(g_{Q}(z_{1}))_{\alpha
^{\prime }\beta ^{\prime }}\}G_{\beta ^{\prime }\beta }^{S}(z_{1}-E_{j})
\notag \\
&&\hspace{-1.5cm}+N^{-1}\sum_{\alpha ^{\prime },\beta ^{\prime },\gamma ^{\prime
},\delta ^{\prime }=1}^{p}\mathbf{E}\{G_{\alpha j,\alpha ^{\prime
}j}(z_{2})Q_{\alpha ^{\prime }\gamma ^{\prime }}F_{\gamma ^{\prime }\delta
^{\prime }\gamma \delta }^{(k)}\}Q_{\delta ^{\prime }\beta ^{\prime
}}G_{\beta ^{\prime }\beta }^{\mathcal{S}}(z_{1}-E_{\substack{ j}})
\label{cga2}
\end{eqnarray}%
Next, we  use (\ref{4g}) and (\ref{vgoc}) to replace $g_{Q}$ and $%
F^{(jk)}$ by their expectations $\overline{g}_{Q}$ and $\overline{F}_{\alpha
\alpha ^{\prime }\gamma \alpha ^{\prime }}^{(jk)}$ in the summand of the
second term of the r.h.s.  yielding%
\begin{equation*}
\overline{F}_{\alpha \alpha ^{\prime }\gamma \alpha ^{\prime
}}^{(jk)}(z_{1},z_{2})(\overline{g}_{Q}(z_{1}))_{\alpha ^{\prime }\beta
^{\prime }}\}G_{\beta ^{\prime }\beta }^{\mathcal{S}}(z_{1}-E_{j})
\end{equation*}%
instead of the term. This allows us to carry out the procedure analogous to that leading from (%
\ref{cgb2}) to (\ref{cgb4}), i.e., replacing $G_{\alpha \beta }^{\mathcal{S}%
}(z_1-E_{j})$ by $G_{\alpha \beta }^{\mathcal{S}}(z_{Q}(z_1)-E_{j})$
and to obtain instead of (\ref{cga2})%
\begin{eqnarray}
\overline{F}_{\alpha \beta \gamma \delta }^{(jk)} &=&\delta _{jk}\overline{G}%
_{\alpha j,\gamma k}(z_{2})G_{\delta \beta }^{\mathcal{S}}(z_{Q}(z_1)-E_{j}) \notag \\
&&\hspace{-2cm}+N^{-1}\sum_{\alpha ^{\prime },\beta ^{\prime },\gamma ^{\prime
},\delta ^{\prime }=1}^{p}\mathbf{E}\{G_{\alpha j,\alpha ^{\prime
}j}(z_{2})Q_{\alpha ^{\prime }\gamma ^{\prime }}F_{\gamma ^{\prime }\delta
^{\prime }\gamma \delta }^{(k)}(z_{1},z_{2})\}Q_{\delta ^{\prime }\beta
^{\prime }}G_{\beta ^{\prime }\beta }^{\mathcal{S}}(z_{Q}(z_1)-E_j).
\label{cga4}
\end{eqnarray}
Next, following the scheme of proof of Result 1, in particular, by using the
relations%
\begin{equation*}
\left\vert \sum_{\alpha \text{ }^{\prime },a=1}^{p,N}G_{\alpha
_{1}j_{1},\alpha ^{\prime }a}(z_{1})G_{\alpha _{2}j_{2},\alpha ^{\prime
}a}^{\ast }(z_{2})=(G(z_{1})G(z_{2}))_{\alpha _{1}j_{1},\alpha
_{2}j_{2}}\right\vert \leq (|\Im z_{1}\Im z_{2}|)^{-1},
\end{equation*}%
instead of (\ref{unit1}), we obtain the bound
\begin{equation}
\mathbf{Var}\{F_{\gamma ^{\prime }\beta ^{\prime }\gamma \delta
}^{(k)}(z_{1},z_{2})\}\leq \frac{2p^{2}||Q||^{2}}{N\eta ^{6}},\;|\Im
z_{1}|,|\Im z_{2}|\geq \eta >0.  \label{varga}
\end{equation}%
The bound allows us to replace $F_{\alpha \alpha ^{\prime }\gamma \alpha
^{\prime }}^{(k)}$ by $\overline{F}_{\alpha \alpha ^{\prime }\gamma \alpha
^{\prime }}^{(k)}=\mathbf{E}\{F_{\alpha \alpha ^{\prime }\gamma \alpha
^{\prime }}^{(k)}\}$ in the second term of the r.h.s. of (\ref{cga4}). In
addition, we will use (\ref{goc}) to replace $\overline{G}_{\alpha j,\beta
k}(z)$ by $\delta _{jk}G_{\alpha \beta }^{\mathcal{S}}(z_Q{z}-E_{j})$
in the first term in \ the r.h.s. of (\ref{cga4}), then we sum the result
over $j=1,...,N$. This converts $\overline{F}^{(jk)}$ into $\overline{F}%
^{(k)}$ in the l.h.s. of (\ref{cgb3}) in view of (\ref{gakpk}) and
\begin{equation*}
N^{-1}\sum_{j=1}^{N}\mathbf{E}\{G_{\alpha j,\alpha ^{\prime
}j}(z_{2})\}G_{\beta ^{\prime }\beta }^{\mathcal{S}}(z_{Q}(z_{1})-E_{j})
\end{equation*}%
into
\begin{equation*}
\int G_{\alpha \alpha ^{\prime }}^{\mathcal{S}}(z_{Q}(z_2)-E)G_{\beta
^{\prime }\beta }^{\mathcal{S}}(z_{Q}(z_1)-E)\nu _{N}(E)dE.
\end{equation*}%
in the second term of the r.h.s. of (\ref{cgb3}) in view of (\ref{non}).
This yields
\begin{eqnarray*}
\overline{F}_{\alpha \beta \gamma \delta }^{(k)} &=&G_{\alpha \gamma }^{%
\mathcal{S}}(z_{Q}(z_2)-E_{k})G_{\delta \beta }^{\mathcal{S}}(z_{Q}(z_1)-E_{k})\\
&&\hspace{-2cm}+\sum_{\alpha ^{\prime },\beta ^{\prime
},\gamma ^{\prime },\delta ^{\prime }=1}^{p}\int G_{\alpha \alpha ^{\prime
}}^{\mathcal{S}}(z_{Q}(z_2)-E)
 Q_{\alpha ^{\prime }\gamma ^{\prime }}F(z_{1},z_{2})Q_{\delta
^{\prime }\beta ^{\prime }}G_{\beta ^{\prime }\beta }^{\mathcal{S}}(z_{Q}(z_1)-E)\nu _{N}(E)dE+R^{(2)},
\end{eqnarray*}%
where $R^{(2)}$ is the sum of error terms resulting from all the replacements
above: $g$ by $\overline{g\text{,}}$ $F^{(k)}$ by $\overline{F}^{(k)}$ and $%
\overline{G}_{\alpha j,\beta k}$ by $\delta _{jk}G_{\alpha \beta }^{\mathcal{%
S}}(\widetilde{z}-E_{k})$. By using an argument similar to that proving (\ref%
{rab2}) and (\ref{cgbf}), it can be shown that the corresponding error terms
are $O(N^{-1})$ provided that $|\Im z_{1,2}|\geq \eta >0$ with an $N$%
-independent $\eta $. This, (\ref{ele}) and (\ref{non}) allow us to carry
out the limit $N\rightarrow \infty $ with (\ref{ele}) in the above relation,
i.e., to show that the limit%
\begin{equation}
F_{\alpha \beta \gamma \delta }(E,z_{1},z_{2})=\lim_{N\rightarrow \infty }%
\overline{F}_{\alpha \beta \gamma \delta }^{(k^{(N)})}(z_{1},z_{2})
\label{flim}
\end{equation}%
exists uniformly in $z_{1,2}$ with $|\Im z_{1,2}|\geq \eta >0$ and satisfies
the equation%
\begin{eqnarray}
F_{\alpha \beta \gamma \delta }(E,z_{1},z_{2}) &=&G_{\alpha \gamma }^{%
\mathcal{S}}(z_{Q}(z_{2})-E)G_{\delta \beta }^{\mathcal{S}}(z_{Q}(z_{1})-E)
\notag\\
&&\hspace{-1.5cm}+\sum_{\alpha ^{\prime },\beta ^{\prime },\gamma ^{\prime },\delta
^{\prime }=1}^{p}\int G_{\alpha \alpha ^{\prime }}^{\mathcal{S}%
s}(z_{Q}(z_{2})-E^{\prime })Q_{\alpha ^{\prime }\gamma ^{\prime }} F_{\alpha \beta \gamma ^{\prime }\delta ^{\prime
}}(E,z_{1},z_{2})\notag\\
&& \times Q_{\delta ^{\prime }\beta ^{\prime }}G_{\beta ^{\prime
}\beta }^{\mathcal{S}}(z_{Q}(z_{1})-E^{\prime })\nu _{0}(E^{\prime
})dE^{\prime }.\label{eqf}
\end{eqnarray}
Multiplying (\ref{eqf}) by $\rho _{\gamma \delta }(0)$ and summing over $%
\gamma $ and $\delta $, we obtain that the $p\times p$ matrix
\begin{equation}
F(E,z_{1},z_{2})=\{F_{\alpha \beta }(E,z_{1},z_{2})\}_{\alpha ,\beta
=1}^{p},\;F_{\alpha \beta }(E,z_{1},z_{2})=\sum_{\gamma ,\delta
=1}^{p}F_{\alpha \beta \gamma \delta }^{{}}(E,z_{1},z_{2})\rho _{\gamma
\delta }(0)  \label{fr0}
\end{equation}
satisfies (\ref{brevG1}) -- (\ref{fo1}).
Applying now to (\ref{brevG1}) the operation defined by (\ref{ilap}) with
respect to the both variables $z_{1}$ and $z_{2}$ and taking into account (%
\ref{gz1}) and (\ref{ligdi}), we obtain finally formulas (\ref{rhosg1}) -- (\ref{neva}) for the limiting reduced
density matrix $\rho (E,z_{1},z_{2})$ defined by (\ref{dmg}) -- (\ref{rdm})
and (\ref{ele}).$\blacksquare$

\medskip
\textbf{Remarks}. (i) Formulas (\ref{gz1}) and (\ref{brevG1}) bear analogy to the well known fact on the mean field approximation in statistical mechanics, where also the first (one-point) correlation function satisfies a nonlinear equation (e.g. the Curie-Weiss equation), while the higher correlation functions are linear in the product of the first correlation function. Analogous situation is in random matrix theory, see e.g. \cite{Mo-Co:92}.

(ii) Consider the case of $p=2$, where $H^{\mathcal{S}}=s\sigma _{z}$ and $Q^{%
\mathcal{S}}=v\sigma _{x}$. In this case $G_{\alpha \beta }^{\mathcal{S}%
}(z)=\delta _{\alpha \beta }r_{\alpha }(z)$,$\;r_{\alpha }(z)=(\alpha
s-z)^{-1}$, $(Q^{\mathcal{S}}G^{\mathcal{S}}(z)Q^{\mathcal{S}})_{\alpha
\beta }=\delta _{\alpha \beta }r_{-\alpha }(z),\;\alpha =\pm $ and we obtain
the basic formulas (4.1) --\ (4.7) of the one-qubit model with random matrix
environment presented and analyzed in \cite{Le-Pa:03}.

\section{Conclusion}
We have considered in this paper the time evolution of quantum correlations of two qubits embedded in a common disordered and multiconnected environment.  We model the environment part of the  corresponding Hamiltonian (\ref{hc1}) by random matrices of large size which can be viewed as a mean field version of the one- (or few-) body Hamiltonians describing complex and not necessarily macroscopic quantum systems.
This continues our study of the two qubit time evolution
carried out in our paper \cite{Br-Pa:18} where the case of two qubits embedded in independent random matrix environments has been studied.

Note that we have used in this paper the Gaussian random matrices (\ref{lGOE}), but our results remain valid for much more general classes of hermitian and real symmetric matrices, in particular, for the so-called Wigner matrices whose entries are independent (modulo the matrix symmetry) random variables satisfying (\ref{GOE}), although in this case the corresponding proofs
are technically more involved, see e.g. Chapter 18 of \cite{Pa-Sh:11} for the corresponding techniques applied to the proof of the Deformed Semicircle Law of random matrix theory.

We have shown that these models are asymptotically exactly solvable in
the limit of large matrix size. By using then an analog of the Bogolyubov - van Hove asymptotic regime, we were able to analyzed a variety of the qubit dynamics ranging between the Markovian (memoryless) and non-Markovian
(including the environment backaction) dynamics.

We have probed the quantum correlation by the widely used numerical characteristics (quantifiers) of quantum states: the negativity, the
concurrence, the quantum discord and the von Neumann entropy.
The first two are sufficiently adequate quantifiers of
entanglement, while the last two quantify also other non-classical
correlations.

For the models with independent environments considered in \cite{Br-Pa:18} the typical behavior of the negativity and the concurrence is the monotone
decay in time from their value at the initial moment to zero at a certain finite moment, the same for the negativity and the concurrence (known as the moment of the so-called Entanglement Sudden Death, ESB). These quantifiers have the qualitatively same behavior for various parameters of the density of states of the environment and entangled initial conditions (being identically zero for the product, i.e., initially unentangled conditions.

For the model with the common random matrix environment of this paper
the situation is quite different because of the indirect interaction
of qubits via the environment. The concurrence
and the negativity for the product states as function of time may
be zero during a certain initial period and become positive
later (the so-called Entanglement Sudden Birth, ESB), may not vanish
at infinity (the so-called entanglement trapping), may have multiple
alternating  ESB's and ESD's and/or damping oscillation.
A strong dependence on the initial conditions and on the density of states of the environment is
also the case.

The behavior of quantum discord proved to be also rather diverse. It may be
zero only at infinity and under special conditions (see Fig. 4b) of the
paper and Fig. 3a) of \cite{Br-Pa:18}). It may attain a finite non zero value
at infinity and may even grow monotonically for large times, may have the plateaux, known as the freezing of the discord \cite{Be-Co:17,Lo-Co:13}, a regular and an oscillating behavior.
Unlike this, the entropy varies regularly in time from zero at the initial moment
to a certain finite value at infinity, see e.g.,  Fig. 4b).

Our results are new in the sense that they are obtained in the framework of a new random matrix model of the qubit evolution which takes into account the dynamical correlations between the qubits via the environment. The results exhibit a variety
of patterns, partly new and partly qualitatively similar to those
found before for the various versions, exact and approximate, of the bosonic environment and can be used in the choice of appropriate models and quantifiers  for quantum information processing with open systems. This can also be
viewed as a manifestation of the universality (the independence on the model) of the patterns, since the environments modeled by free boson field and by random matrices of large size correspond to
seemingly different physical situations.


\end{document}